\documentclass[hyper,12pt,letterpaper]{JHEP3}
\usepackage{amsmath,amssymb,multirow,array}

\newcommand{\f}{\frac}
\newcommand{\p}{\partial}

\newcommand{\s}{\sigma}
\newcommand{\half}{\frac{1}{2}}

\newcommand{\cd}{\!\cdot\!}
\newcommand{\str}{{*}}

\newcommand{\sthr}{\sigma_3} 
\newcommand{\sfou}{\sigma_4} 
\newcommand{\sfiv}{\sigma_5} 
\newcommand{\seig}{\sigma_8} 
\newcommand{\snin}{\sigma_9} 
\newcommand{\sten}{\sigma_{10}} 

 \newcommand{\hT}{\hat T}
 \newcommand{\hn}{\hat \nu}
 
\newcommand{\hm}{\hat \mu}

\numberwithin{equation}{section}

\newcommand{\nn}{\nonumber}

\title{Constraints on Superfluid Hydrodynamics from Equilibrium Partition
Functions}
\author{Sayantani Bhattacharyya$^a$,
Sachin Jain$^b$, Shiraz Minwalla$^b$ and Tarun Sharma$^b$\\
$^a$Harish-Chandra Research Institute, Chhatnag Road, Jhunsi, Allahabad-211019.\\
$^b$Dept. of Theoretical Physics, Tata Institute of Fundamental Research, Homi Bhabha Rd,
Mumbai 400005, India. \\
Email:\ \ {\bf sayanta@hri.res.in,
minwalla@theory.tifr.res.in, sachin@theory.tifr.res.in, tarun@theory.tifr.res.in
}}

\abstract{Following up on recent work in the context of ordinary fluids, we 
study the equilibrium partition function of a 3+1 dimensional 
superfluid on an arbitrary stationary 
background spacetime, and with arbitrary stationary background gauge fields, 
in the long wavelength expansion. We argue that this partition function is 
generated by a 3 dimensional Euclidean effective action for the massless 
Goldstone field. We parameterize the general form of this action 
at first order in the derivative expansion. We demonstrate that the
constitutive relations of relativistic superfluid hydrodynamics 
are significantly constrained by the requirement 
of consistency with such an effective action. At first order in the 
derivative expansion we demonstrate that the resultant 
constraints on constitutive relations coincide precisely with 
the equalities between hydrodynamical transport coefficients 
recently derived from the second law of thermodynamics.}

\preprint{TFR/TH/12-26}

\begin{document}
\maketitle

\section{Introduction}

Hydrodynamics is the long distance effective description 
of locally thermalized systems. The variables
of hydrodynamics are local values of the temperature, chemical potential, 
velocity and other relevant thermodynamical order parameters. 
The equations of hydrodynamics are the universal laws of 
conservation of the stress tensor and the charge current. 
These equations may be used to describe the propagation of a fluid 
in an arbitrary weakly curved background metric, and with arbitrary 
slowly varying background gauge fields. Within the hydrodynamical 
approximation, the stress tensor and charge current of the fluid 
are expressed as functions of 
thermodynamical and background fields; the formulas through which 
this is achieved are referred to as the constitutive relations of the 
hydrodynamical system. 
 
The hydrodynamical constitutive relations of any given system 
may, in principle, be determined by a detailed study of the dynamics of the 
theory. For strongly coupled quantum field theories, however, the required 
calculations usually cannot be practically executed. 
\footnote{The only exceptions that we are aware of 
lie within the fluid gravity map ( \cite{Bhattacharyya:2008jc}, see 
see \cite{Rangamani:2009xk,Hubeny:2011hd} for reviews) 
of the AdS/CFT correspondence of string 
theory. The constitutive relations of field theories with a known 
dual description are rather easily determined following the procedure 
first described in \cite{Bhattacharyya:2008jc}.} Given this state of affairs, it is clearly of 
interest to have a complete parameterization of the most general hydrodynamical 
constitutive relations allowed on general grounds. Such a characterization 
would constitute a satisfactory framework for the the theory classical 
hydrodynamics viewed as an autonomous long wavelength effective theory.

The constitutive relations of relativistic 
hydrodynamics are specified in an expansion
in derivatives of the local thermodynamical fields and background fields. 
At any order in this 
expansion, Lorentz invariance determines the constitutive relations 
up to a finite number of functions of the scalar thermodynamical fields 
(e.g. local temperature and chemical potential). It turns out, however, 
that other considerations further constrain the constitutive relations. 
These constraints are of two sorts. The first, and more important, set 
of constraints asserts relations between the apparently independent
functions that appear in the most general Lorentz allowed constitutive 
relations. Such constraints cut down the number of free functions 
in the equations of hydrodynamics (at any order in the derivative 
expansion). We refer to constraints of this first sort as 
`equality type' constraints; they are the primary focus of the current 
paper. A second, milder form of constraints assert inequalities for 
the free functions that appear in constitutive relations. These constraints
do not reduce the number of free functions in constitutive relations but
merely bound these functions. In this paper we will have nothing to say 
about this second class of constraints.

One method for obtaining constraints on constitutive relations was outlined
in the classic text book of Landau and Lifshitz \cite{Landau:1987gn} and 
refined in later studies ( see e.g. \cite{Son:2009tf, Bhattacharya:2011tra, 
Lin:2011aa, Neiman:2011mj, Jensen:2011xb, Bhattacharyya:2012ex} for recent 
work). 
This method is based on the assumption that 
consistent equations of hydrodynamics come equipped with an 
entropy current.  Like the conserved currents, the entropy current is  
a function of the local thermodynamical fields. The key dynamical assumption 
is that the divergence of this entropy current is point 
wise (in spacetime) positive semi definite for every conceivable fluid flow. 
We refer to the existence of such a positive divergence entropy current - 
without making any prior assumptions as to its functional form -  as
a local version of the second law of thermodynamics. The local second 
law clearly guarantees that net thermodynamical entropy increases in any 
 fluid flow that starts and ends in equilibrium. 
The requirement that entropy increase for fluid flows perturbed by 
arbitrary local 
sources suggests that the local form of the second law is also a necessary 
consequence of the second law. \footnote{It is at least suggestive 
that the fluid dynamics generated by the 
fluid gravity map of the AdS/CFT correspondence does indeed always 
obey the local form of the second law, at least when the dual bulk description
is given by the equations of two derivative Einstein gravity with matter 
that obeys the null energy conditions.}

At low orders in the derivative expansion it has been demonstrated 
that the local version of the second law of thermodynamics yields 
powerful  constraints (of both the equality and inequality sort) on 
otherwise unrelated free functions in constitutive relations. The detailed 
form of these 
constraints has been worked out for uncharged relativistic 
fluids up to second order in the derivative expansion 
\cite{Bhattacharyya:2012ex}  and for (parity non preserving) 
charged fluids to first order in the derivative expansion (see  
\cite{Son:2009tf, Neiman:2010zi, Bhattacharya:2011tra} for 3+1 dimensional fluids and
\cite{Jensen:2011xb} for 2+1 dimensional fluids).

Very recently, a second systematic 
method for constraining the constitutive relations for fluids was 
described in \cite{Banerjee:2012iz,Jensen:2012jh}\footnote{See also 
\cite{Loganayagam:2011mu,Loganayagam:2012pz} for earlier related work, \cite{Jensen:2012jy} for closely 
related discussions and \cite{Jain:2012rh, Valle:2012em} for follow 
up work.}.
 These constraints follow from the very reasonable 
demand that the hydrodynamical equations must always admit 
equilibrium solutions for arbitrary stationary background field 
configurations, and moreover that the values of conserved charges in 
equilibrium must be consistent with the existence of an equilibrium partition 
function. For both uncharged fluids at second order as well as charged 
parity non invariant (and potentially anomalous) fluids at first order, the 
equality type constraints obtained from this method agree, 
in full detail \cite{Banerjee:2012iz,Jensen:2012jh} with those obtained from 
the local form of the 
second law of thermodynamics described above. It has been 
conjectured \cite{Banerjee:2012iz} 
that this agreement persists to all orders in the 
derivative expansion; however there is as yet no proof of this 
conjecture. 

The equations of charged hydrodynamics are modified when the 
charge symmetry of the system is spontaneously broken 
by the condensation of a charged operator in thermal equilibrium.
The effective description of such systems has new hydrodynamical 
degrees of freedom whose origin lies in the Goldstone mode of the 
charge condensate. The resultant hydrodynamical equations are 
referred to as the equations of superfluid hydrodynamics, and 
are the subject of the current paper. 

More particularly in this paper we study 
`s' wave superfluid hydrodynamics, i.e. the hydrodynamics of a 
system whose charge condensate is a complex scalar operator. 
We study the constraints on the equations of 
first order `s' wave superfluid hydrodynamics imposed by the requirement that 
these equations admit equilibrium under appropriate situations, 
and that the charge currents in equilibrium agree with those
from an appropriate partition function. We do not assume that 
the superfluids we study necessarily preserve either parity 
or time reversal invariance. 

As we explain in section \ref{epf} below, the general analysis presented 
in this paper closely follows that of \cite{Banerjee:2012iz} (for the case of ordinary, i.e. 
not `super' fluids) with one important difference. The Euclidean 
partition function for a superfluid in an arbitrary background 
\footnote{See \cite{Jensen:2012jh,Dubovsky:2011sj} for a discussion of this partition function
at the perfect fluid level.} is 
determined by an effective  field theory that includes a massless mode: 
the Goldstone boson of the theory. This effective field theory 
is local, and may usefully be studied in the derivative expansion.
However the partition function that follows after integrating out the 
Goldstone boson is neither local nor simple. As we explain below, the 
study of the local effective action of the Goldstone boson (rather than the 
partition function itself) allows us to usefully constrain
the constitutive relations of superfluid hydrodynamics. In this paper 
we present a careful derivation of the relations between otherwise 
independent transport functions that follow from such a study.

Constraints on the constitutive relations of first order superfluid 
hydrodynamics have previously been obtained using the local form of the 
second law in \cite{putterman, Lin:2011aa,Bhattacharya:2011tra,Neiman:2011mj} for the case of time reversal invariant superfluids. 
In this paper we generalize the derivation of \cite{Bhattacharya:2011tra} to include the study 
of superfluids that do not preserve time reversal invariance.  We then compare
the results obtained from the two different methods; i.e. the 
constraints that follow from the requirement of existence of 
equilibrium and those that follow from the local second law. As in the   
case of ordinary (i.e. non super) fluids we find perfect agreement 
between the equality type constraints obtained from these two 
apparently distinct methods. Our results supply further evidence for the 
conjecture that the equality type constraints from these two methods 
agree in a wide range of hydrodynamical 
contexts and to all orders in the derivative expansion. 
 A proof of this conjecture would go some way towards 
proving the local form of the second law, and would permit the demystification
of this law in a hydrodynamical context. 

While the work reported in this paper is purely hydrodynamical and nowhere 
uses AdS/CFT, much of the motivation for this work lies within the 
fluid gravity map of AdS/CFT. The status of the second law of thermodynamics 
for theories of gravity that include  higher derivative corrections to 
the Einstein Lagrangian remains unclear. In particular 
it has never been proved that the Hawking area increase theorem generalizes 
to a Wald entropy increase theorem for arbitrary higher derivative corrections
to Einstein's gravity. If the interplay between the existence of equilibrium 
in appropriate circumstances and entropy increase can be proved on general 
grounds in a hydrodynamical context, then it seems likely that the lessons 
learnt can be taken over to the study of entropy increase in higher derivative
gravity (at least for asymptotically AdS space) via the fluid gravity map. 
This could lead to a proof of a Wald entropy increase theorem under 
appropriate conditions on the higher derivative corrections of the
gravitational system.

\section{Equilibrium effective action for the Goldstone mode}\label{epf}

\subsection{The question addressed}

In this section we study an $s$ wave superfluid propagating on the stationary 
background metric
\begin{equation} \label{metgf}
ds^2=G_{\mu \nu} dx^\mu dx^\nu 
=-e^{2 \sigma(\vec x)} \left( dt+ a_{i}(\vec x) dx^i \right)^2
+g_{ij}({\vec x}) dx^i dx^j
\end{equation}
and background gauge field 
\begin{equation} \label{gf}
A= {\cal A}_0(\vec x) dx^0+ {\cal A}_i (\vec x) dx^i
\end{equation} 
Below we will often work in terms of the modified gauge fields 
\begin{equation}\label{sgf} \begin{split}
&A_i={\cal A}_i -A_0 a_i \\
&A_0={\cal A}_0 +\mu_0
\end{split}
\end{equation}
All background fields above are assumed to vary slowly; we work
in an expansion in derivatives of these fields.  We address the following question: what is the most 
general allowed form of the partition function
\begin{equation}\label{pftrd}
Z={\rm Tr} e^{-\frac{H -\mu_0 Q}{T_0}}
\end{equation}
as a function of the background fields 
 $\sigma$, $a_i$, $g_{ij}$, $A_0$ and $A_i$ in a systematic derivative 
expansion?

\subsection{The partition function for charged (non super) fluids}

The analogous question was studied for the case of an 
ordinary (non super) charged fluid in \cite{Banerjee:2012iz}. It was demonstrated that 
to first order in the derivative expansion the most general allowed form of 
the partition function for an ordinary charged fluid on the background 
\eqref{metgf}, \eqref{gf} is given by 
\begin{equation}\label{cfacn} \begin{split}
W&=\ln Z = W^0+W^1_{inv} +W^1_{anom}\\
W^0 &= \int \sqrt{g} \frac{e^{\sigma}}{T_0} P\left(T_0 e^{-\sigma}, e^{-\sigma} 
A_0 \right) \\
W^1_{inv}&= \frac{C_0}{2}  \int A d A  
+ \frac{C_1}{2} \int  a d a + \frac{C_2}{2} \int A d a   \\
W^1_{anom}&= \frac{C}{2} \left( \int \frac{A_0}{3} A d A 
+ \frac{A_0^2}{6} A d a \right)
\end{split}
\end{equation}
where $P(T, \mu)$ is the thermodynamical pressure of the system as a 
function of its temperature and chemical potential  and $C_0$, $C_1$ $C_2$ and 
$C$ are all constants. The constant $C$ specifies the 
covariant $U(1)^3$ anomaly via the equation
\begin{equation}\label{gie}
\partial_\mu {\tilde J}^\mu = -\frac{C}{8} *(F \wedge F)
\end{equation} 
The constants $C_0$, $C_1$ and $C_2$ do not (yet) have similar interpretations. 
It was demonstrated that $C_0=C_1=0$ in any system that respects CPT 
invariance.

Notice that the result \eqref{cfacn} for the partition function of 
an ordinary (non super) fluid is a {\it local } function of the 
background sources $g_{ij}$, $a_i$, $\sigma$, $A_0$ and $A_i$. 
Locality is a direct 
consequence of the fact that the path integral that computes 
the partition function \eqref{pftrd} has a unique hydrodynamical saddle point 
(as opposed to a moduli space of saddle points). As a consequence  
the partition function 
 is generically \footnote{Non hydrodynamical massless modes 
occur when the system is tuned to a second order phase transition. We 
assume in what follows that our system has not been tuned to such a phase 
transition. We leave the study of this interesting special case \cite{RevModPhys.49.435} 
to future work.} computed by a path integral over an action with 
no massless fields. It follows that the result is  
local on length scales large compared to the inverse mass gap in the 
action (this mass gap is sometimes referred to as a static screening 
length of the 4 d thermal system)\footnote{We thank K. Jensen for discussions on this topic}. 

\subsection{Euclidean action for the Goldstone mode for superfluids}

Unlike an ordinary charged fluid, the equilibrium configuration 
of a superfluid in the background \eqref{metgf} is not unique. As superfluids
break the global $U(1)$ symmetry, every background admits at least a one 
parameter set of equilibrium configurations 
that differ by a constant shift in the phase of the expectation value of the 
condensed scalar. It follows that the path integral that computes 
\eqref{pftrd} has a zero mode (the phase of the scalar 
condensate). Consequently, the partition function 
\eqref{pftrd}, is {\it not} a local function of the background source 
fields. Instead this partition function is generated by a local three 
dimensional field theory of the {\it dynamical} phase field $\phi$. 

The dynamics of the Goldstone boson in general,  governed by a 3d massless
quantum field theory. In this paper, however, we focus on field theories 
in an appropriate large N limit (such as theories with matrix degrees 
of freedom in the t' Hooft limit). In such a limit the effective action 
for the Goldstone boson is multiplied by a suitable positive power of 
$N$ (the factor is $N^2$ in the t'Hooft limit mentioned above). As a 
consequence Goldstone dynamics is effectively classical in the large 
$N$ limit. Quantum corrections to this classical answer, which are suppressed 
by appropriate powers of $N$ (this power is $\frac{1}{N^2}$ in the t'Hooft
limit), may have very interesting structure, see e.g. \cite{Kovtun:2003vj,CaronHuot:2009iq,Kovtun:2011np,Kovtun:2012rj} for related 
work. We leave their 
study to future work. \footnote{We thank K. Jensen for discussions on this 
topic.}

In principle, the partition function \eqref{pftrd} for the superfluid may be 
obtained  from the corresponding local effective action by integrating out the 
Goldstone boson (i.e. solving its equation of motion and plugging the solution
back into the action). \footnote{If the Euclidean 3 dimensional manifold 
we work on is compact and we demand single valuedness of the field $\phi$ 
then it is plausible that the solution to the $\phi$ equation of motion 
is (at least generically) unique, see below.}
In practice the implementation of this procedure 
requires the solution of a nonlinear partial differential equation. Moreover, 
even if one could solve this equation the resultant 
partition function would be highly nonlocal. A direct analysis of the 
partition function itself seems neither easy nor particularly useful. 
In order to obtain constraints on the equations 
of superfluid hydrodynamics below we will work directly with the 
local effective action for the Goldstone mode rather than the final 
result for the partition function. 

The requirements of gauge invariance significantly constrain the form of 
Goldstone effective action. Let $\phi$ 
denote the phase of the scalar condensate. Under a gauge transformation 
${\cal A}_i  \rightarrow {\cal A}_i + \partial_i \alpha$, $\phi$ 
transforms as 
$\phi+\alpha$. It follows that the effective action 
can only depend on the combination 
$$\xi_i=-\partial_i \phi + {\cal A}_i$$
Note that $\xi_\mu$ like ${\cal A}_\mu$, is a field of zero order in the 
derivative 
expansion \footnote{This means that the phase field $\phi$ is of $-1$ 
order in derivatives; this observation does not invalidate the 
derivative expansion  as gauge invariant physical quantities are 
functions only of $\xi^\mu$ and not independently 
of $\phi$.}.

The local field theory for the Goldstone boson must also enjoy invariance 
under  Kaluza Klein gauge transformations
($a_i \rightarrow a_i - \partial_i \gamma$, see subsection $2.2$ of \cite{Banerjee:2012iz} for 
details). For this reason we work with the Kaluza Klein invariant fields 
\begin{equation}
{\zeta}_{i}=\xi_{i}-a_{i} A_{0}=-\partial_i \phi + A_i.\\
\end{equation}
We also define 
$$\xi_0=A_0$$
and define 
\begin{equation} \label{defchi}
\chi =  \xi^{2}=-\xi_{\mu}\xi^{\mu}=\xi_{0}^{2}e^{-2\sigma}-g^{ij}{\zeta}_{i}
\zeta_{j}.
\end{equation}

\subsection{The Goldstone action for perfect superfluid hydrodynamics}

As we have explained above, the euclidean partition function for our system 
is generated by an effective action $S$ for the Goldstone field $\phi$.
This Goldstone action may be expanded in a power series in derivatives.
\begin{equation} \label{actexp}
S=S_0+S_1+S_2 \ldots 
\end{equation}
At lowest (zero) order in the derivative expansion symmetries constrain the 
Goldstone boson effective action to take the form\footnote{The action
\eqref{zopf} was already presented in \cite{Jensen:2012jh}. The presentation of 
this subsection differs from \cite{Jensen:2012jh} only in the emphasis that 
$\phi$ be regarded as a dynamical field in \eqref{zopf}, rather than a 
background like ${\hat T}$. For related discussions on effective action for superfluid, see for example \cite{Son:2002zn,Dubovsky:2011sj}.}
\begin{equation}\label{zopf} \begin{split}
 &S_0= \int d^{3}x\sqrt{g}\frac{1}{{\hat T}} P({\hat T},{\hat \mu},\chi).\\
& {\hat T}= T_0 e^{-\sigma} \\
& {\hat \mu}=A_0 e^{-\sigma}\\
& {\hat u}^\mu=(1,0,0,0) e^{-\sigma}\\
\end{split}
\end{equation}
where $P$ is an arbitrary function whose thermodynamical significance we 
will soon discover, and $\chi$ was defined in \eqref{defchi}.
The fields ${\hat T}$, ${\hat \mu}$ and ${\hat u}^\mu$ are the values 
of the hydrodynamical temperature, chemical potential and velocity  
fields in equilibrium at zeroth order in the derivative expansion
(see \cite{Banerjee:2012iz}).

In the classical (or large $N$) limit adopted throughout this paper, the 
partition function $Z$ of our system is obtained by evaluating the 
Goldstone action on shell. Let the solution to the 
equation of motion be denoted by 
$$\zeta_i(x)=\zeta^{eq}_i(x).$$ 
Then the partition function is given by
\begin{equation} 
\ln Z = S(\zeta^{eq}_i(x))
\end{equation}

At lowest order in the derivative expansion, the action \eqref{zopf} depends 
only on first 
derivatives of the massless field $\phi$. Varying this action w.r.t. 
$\phi$ 
\begin{eqnarray} \label{pfi}
 \delta S_0&=&\int d^{3}x\sqrt{g}\frac{e^{\sigma}}{T_{0}} 
\frac{\partial P}{\partial \chi} 2 g^{ij}\zeta_{i}\partial_{j}\delta \phi\nn\\
&=&-\int d^{3}x \frac{1}{T_0} \partial_{j}(\sqrt{-G}f \zeta^{j})\delta\phi
\end{eqnarray}
yields 
\begin{equation}\label{eqxi}
  \partial_{j}(\sqrt{-G}f \zeta^{j})=\nabla^{(4)}_{\mu}(f \xi^{\mu}) = \nabla_i \left(\frac{f}{T}\zeta^i\right)=0.
\end{equation}
where
$$ f= 2\frac{\partial P}{\partial \chi}.$$
Note this equation of motion is of second order in derivatives of the field 
$\phi$. 
\footnote{The formal similarity of 
\eqref{eqxi} to the equation $\nabla^2 \phi=0$ 
(where the Laplacian is taken in an appropriately rescaled metric) suggests 
that \eqref{eqxi} has a unique solution on a compact manifold (up to constant 
shift in $\phi$) provided that $\phi$ is required to be single valued 
and smooth on this manifold. However we do not have a proof of this statement.} Plugging the solution to \eqref{eqxi}  back into the 
\eqref{pfi} in principle yields an explicit though complicated and nonlocal 
 expression for the partition function of the system as a function of 
source fields. 

The stress tensor and charge current that follow from the action 
\eqref{zopf} may be computed in a straightforward manner using the formulas 
listed in eqs.($2.16$) of \cite{Banerjee:2012iz}; they are given by 
\begin{equation}\begin{split}\label{cc} 
J_0 &= -\frac{T_0 e^{\sigma}}{\sqrt{g}} \frac{\delta S_0}{\delta A_0} = -e^{2\sigma} \left[ e^{-\sigma} \frac{\partial P}{\partial \mu} + 
\frac{\partial P}{\partial \chi} \frac{\partial \chi}{\partial A_0} \right] = -qe^\sigma - \xi_0 f \\
J^i &= \frac{T_0 e^{-\sigma}}{\sqrt{g}} \frac{\delta S_0}{\delta A_i} = \frac{\partial P}{\partial \chi} \frac{\partial \chi}{\partial A_i} =- f \xi^i \\
T_{00} &= -\frac{T_0 e^{\sigma}}{\sqrt{g}} \frac{\delta S_0}{\delta \sigma} 
        = -e^{2\sigma} \left[ P + \frac{\partial P}{\partial T_0 e^{-\sigma}} \frac{\partial T_0 e^{-\sigma}}{\partial \sigma} 
          + \frac{\partial P}{\partial \mu} \frac{\partial \mu}{\partial \sigma} 
          + \frac{\partial P}{\partial \chi} \frac{\partial \chi}{\partial \sigma} \right] \\
&=-e^{2\sigma} \left[P-s T-q \mu-f \xi_{0}^2 e^{-2\sigma}\right]=e^{2\sigma} \epsilon+ f \xi_{0}^2\\
T_0^i &= \frac{T_0}{e^{\sigma} \sqrt{g}} \left[ \frac{\delta S_0}{\delta a_i} - A_0 \frac{\delta S_0}{\delta A_i} \right] 
       = \frac{\partial P}{\partial a_i} - A_0 \frac{\partial P}{\partial A_i} = -A_0\frac{\partial P}{\partial \chi} \frac{\partial \chi}{\partial A_i}  =  f A_0 \xi^i \\
T^{ij} &= \frac{-2T_0}{e^\sigma \sqrt{g}} g^{ik}g^{jl} \frac{\delta S_0}{\delta g^{kl}} 
        = -2 g^{ik}g^{jl} \left[ -\half g_{kl} P +  \frac{\partial P}{\partial \chi} \frac{\partial \chi}{\partial g^{kl}} \right] 
        = P g^{ij} + f \xi^i \xi^j \\
\end{split}
\end{equation}
The gauge and diffeomorphism invariance of the action \eqref{zopf} 
ensure the stress tensor and charge current described above are 
automatically conserved onshell (i.e. upon imposing the equation of motion 
\eqref{eqxi}). 

The complicated looking expressions for the conserved currents 
\eqref{cc} may actually be summarized in a remarkably simple form as 
\begin{eqnarray} \label{lto}
 T^{\mu\nu} &=& (\epsilon+ P){\hat u}^{\mu}{\hat u}^{\nu}+ P g^{\mu\nu}+ 
f \xi^{\mu}\xi^{\nu}\nn\\
J^{\mu}&=&q {\hat u}^{\mu}-f \xi^{\mu},
\end{eqnarray}
where ${\hat u}$ was defined in \eqref{zopf} and 
all terms on the RHS of \eqref{lto} are evaluated on the zero order 
equilibrium solutions $T(x)={\hat T}$ and $\mu(x)={\hat \mu}$, defined in 
\eqref{zopf} and the functions $\epsilon$, $s$ and $q$ are defined 
in terms of the pressure $p$ by the equations
\begin{eqnarray} \label{ltt}
\epsilon+ P &=& sT +q    \mu\nn\\
dP&=& s dT+ qd\mu+\frac{1}{2}f d\chi
\end{eqnarray} 
\eqref{lto} and \eqref{ltt} are precisely the Landau-Tisza constitutive 
relations of superfluid hydrodynamics.

\subsection{The Goldstone Action at first order in derivatives}

One derivative corrections to the Goldstone action \eqref{zopf} may be 
divided into parity even and parity odd terms. We 
consider these in turn. 

\subsubsection{Parity even one derivative corrections}\label{sec:parityeven}

The most general parity preserving one derivative correction to 
\eqref{zopf} is given by  
\begin{equation} \begin{split}\label{ppa}
S&=S_0+ S_1^{even}\\
 S_1^{even} &=  \int d^3y \sqrt{g} \left[\frac{f_1}{\hat T}(\zeta.\partial)\hat T 
+\frac{f_2}{\hT}(\zeta.\partial)\hat \nu - 
f_3 \nabla_i\left(\frac{f}{\hat T} \zeta^i\right)\right]
\end{split}
\end{equation}
where ${\hat T}$ was defined in \eqref{zopf},  
$$\hat\nu = \frac{\hat \mu} {\hat T}=  \frac{A_0}{T_0}$$
and 
$$f_i= f_i(\hat T,\hat \nu, \zeta^2)~~~(i = 1 \ldots 3) $$
are arbitrary functions while 
$f$ was defined in the previous subsection
$$f(\hat T,\hat \nu, \zeta^2) = - 2 \frac{\partial P}{\partial \zeta^2}$$

Two remarks are in order 
\begin{itemize}
\item{1.} In \eqref{ppa} the unspecified function $f_3$ multiplies the 
zero order equation of motion of the phase field $\phi$. As a consequence, 
under the field redefinition
\begin{equation}\label{fr} \begin{split}
 \phi &=\tilde\phi + \delta\phi(\hT, \hn, \zeta)\\
\Rightarrow \xi_\mu&  = \tilde\xi_\mu  - \partial_\mu \left( \delta\phi \right)
\end{split}
\end{equation}
we find 
\begin{equation}\label{shift1}
 S_0[\phi] = S_0[\tilde\phi]  -
\int d^{3}x \sqrt{g} \nabla_{j}\left(\frac{f}{\hT} 
\zeta^{j}\right) \delta\phi
\end{equation}
In other words we are free to use the variable $\tilde \phi$ instead of 
$\phi$; however the first derivative correction with this choice 
of variable, $\tilde S_1^{even}$, differs from $S_1^{even}$ by 
\begin{equation}\label{diffts}
\tilde S_1^{even} =S_1^{even} -
\int d^{3}x \sqrt{g} \nabla_{j}\left(\frac{f}{\hT} 
\zeta^{j}\right) \delta\phi
\end{equation}
In other words the field redefinition \eqref{fr} induces the shifts
\begin{equation}\label{shift2}
\tilde {f}_1-f_1 = 0,~~ {\tilde f}_2-f_2= 0 ,~~ {\tilde f}_3-f_3=
\delta\phi
\end{equation}
(where ${\tilde f}_1$, ${\tilde f}_2$ and ${\tilde f}_3$ are the functions 
that appear in the expansion of ${\tilde S}_1^{even}$, see \eqref{ppa} )
For this reason, the dependence of all physical quantities - like the fluid 
constitutive relations - on $f_3$ is rather trivial, and easy to deduce on 
general grounds, as we will see below.
\item{2.}
While the fields $\sigma$, $\mu$ and $\chi$ are even 
under the action of time reversal, the fields ${\xi}_i$ and 
$\zeta_i$  are odd 
under this operation. It follows that each of the three terms in \eqref{fr}
is odd under the action of time reversal. In other words the simultaneous
requirement of parity and time reversal invariance simply sets $W_1=0$. 
It follows that time reversal invariant superfluids have no 
non dissipative
transport coefficients at first order. 
\end{itemize}

%

The corrections from \eqref{zopf} to the  charge current and stress tensor  \eqref{cc} 
in equilibrium are given by 
\begin{equation}\label{evencurrent}
\begin{split}
\delta J_0 &=-\frac{\hT e^{2\sigma}}{\sqrt g}\left[\frac{\delta {S_1^{even}}}{\delta A_0}\right]_{\zeta= \zeta^{eq}}= -\frac{\hT e^{2\sigma}}{\sqrt g}\left(\frac{\delta {W_1^{even}}}{\delta A_0}\right)= -\frac{e^{\sigma}}{\sqrt g}\left(\frac{\delta {W_1^{even}}}{\delta \hn}\right)\\
&=- e^{\sigma}\left[\frac{\partial}{\partial \hn}\left(\frac{f_1}{\hT}\right)(\zeta^{eq}.\partial)\hT + \frac{\partial}{\partial \hn}\left(\frac{f_2}{\hT}\right)(\zeta^{eq}.\partial)\hn + \frac{\partial}{\partial \hn}\left(\frac{f}{\hT}\right)(\zeta^{eq}.\partial)f_3 - \frac{f}{\hT} (\zeta^{eq}.\partial)\left(\frac{f_2}{f}\right)\right]\\
\\
\delta J^i &=\frac{\hT}{\sqrt g}\left( \frac{\delta {S_1^{even}}}{\delta A_i}\right)_{\zeta= \zeta^{eq}}= \frac{\hT}{\sqrt g}\left( \frac{\delta {W_1^{even}}}{\delta A_i}\right)\\
&=2(\zeta^{eq})^i\left[\frac{\partial f_1}{\partial (\zeta^{eq})^2}(\zeta^{eq}.\partial)\hT + \frac{\partial f_2}{\partial (\zeta^{eq})^2}(\zeta^{eq}.\partial)\hn + \frac{\partial f}{\partial (\zeta^{eq})^2}(\zeta^{eq}.\partial)f_3 \right]\\
 &~~+g^{ij}\left(f_1\partial_j \hT +f_2\partial_j \hn +f\partial_j f_3\right)\\
 \end{split}
 \end{equation}
 \begin{eqnarray}\label{evenstress}
  \delta T_{00} &=&  -\frac{\hT e^{2\sigma}}{\sqrt g}\left[\frac{\delta {S_1^{even}}}{\delta \sigma}\right]_{\zeta=\zeta^{eq}} = -\frac{\hT e^{2\sigma}}{\sqrt g}\left(\frac{\delta {W_1^{even}}}{\delta \sigma}\right)= \frac{\hT^2 e^{2\sigma}}{\sqrt g}\left(\frac{\delta {W_1^{even}}}{\delta \hT}\right)\nn\\
&=& \hT^2 e^{2\sigma}\Bigg[\frac{\partial}{\partial \hT}\left(\frac{f_1}{\hT}\right)(\zeta^{eq}.\partial)\hT + \frac{\partial}{\partial \hT}\left(\frac{f_2}{\hT}\right)(\zeta^{eq}.\partial)\hn \nn\\
&&~~~~+ \frac{\partial}{\partial \hT}\left(\frac{f}{\hT}\right)(\zeta^{eq}.\partial)f_3 - \frac{f}{\hT} (\zeta^{eq}.\partial)\left(\frac{f_1}{f}\right)\Bigg]\nn\\
\\
\delta T^i_0 &=& \frac{\hT}{\sqrt g}\left( \frac{\delta {S_1^{even}}}{\delta a_i}\right)|_{\zeta = \zeta^{eq}}=\frac{\hT}{\sqrt g}\left( \frac{\delta {W_1^{even}}}{\delta a_i}\right)|_{A_i= Constant} = -A_0 ~\delta J^i \nn\\
\\
 \delta T^{ij} &=&- \frac{\hT}{\sqrt g}g^{il}g^{jm}\left[\frac{\delta {S_1^{even}}}{\delta g^{ij}}\right]_{\zeta=\zeta^{eq}}=- \frac{\hT}{\sqrt g}g^{il}g^{jm}\left(\frac{\delta {W_1^{even}}}{\delta g^{ij}}\right)\nn\\
 &=& -\left[(\zeta^{eq})^i \delta J^j +(\zeta^{eq})^j \delta J^i\right] +g^{ij}\left[f_1(\zeta^{eq}.\partial)\hT + f_2(\zeta^{eq}.\partial)\hn + f(\zeta^{eq}.\partial)f_3\right]\nn\\
 &&+2(\zeta^{eq})^i(\zeta^{eq})^j\Big[\frac{\partial f_1}{\partial (\zeta^{eq})^2}(\zeta^{eq}.\partial)\hT + \frac{\partial f_2}{\partial (\zeta^{eq})^2}(\zeta^{eq}.\partial)\hn 
+\frac{\partial f}{\partial (\zeta^{eq})^2}(\zeta^{eq}.\partial)f_3 \Big]
\end{eqnarray}
In equations \eqref{evencurrent} and \eqref{evenstress} all the scalar functions $f_1, f_2, f_3$ and $f$ have been treated as functions $\hT, \hn$ and $(\zeta^{eq})^2$ respectively. In obtaining \eqref{evencurrent} 
 we have used the zeroth order equation of motion for $\phi$.
 $$\nabla_i\left(\frac{f}{\hT}(\zeta^{eq})^i\right)=0$$
to simplify the expressions presented above .

\subsubsection{Parity violating terms}

The most general parity odd contributions to the action are given by\footnote{Our convention is $\frac{1}{2}\int XdY =\int d^3 x \sqrt{g_3} \epsilon^{ijk}X_i \partial_{j}Y_k.$} 
\begin{equation}\begin{split}\label{poi}
S^{odd}&=S^{odd}_1+S_{anom}\\
 S^{odd}_1&=\int \sqrt{g}d^{3}x\left(g_1 \epsilon^{ijk}\zeta_{i}\partial_{j}A_k+T_0 g_2 \epsilon^{ijk}\zeta_{i}\partial_{j}a_k \right)  + \frac{C_1}{2} \int a d a \\
S_{anom}&=\frac{C}{2} \left( \int \frac{A_0}{3} A d A 
+ \frac{A_0^2}{6} A d a \right)\\
\end{split}\end{equation}
 \footnote{The action for parity odd (non super) fluids \eqref{cfacn}
also contains the terms 
$$W=\frac{C_0}{2}  \int A d A  
+ \frac{C_2}{2} \int A d a .$$ But using the fact that $\zeta_{i}=A_i+\p_i \phi$ and $$\int \sqrt{g}\epsilon^{ijk}\p_i \phi \p_j A_{k}=0,$$ 
we can absorb $C_0$ in $g_1$ and $C_2$ in $g_2.$}
where
$$g_1=g_1({\hat T}, {\hat \nu}, \psi), ~~~g_2=g_2({\hat T}, {\hat \nu}, \psi),$$
$C_1$ is a constant and 
$${\hat \nu}=\frac{\hat \mu}{\hat T}, ~~~\psi=\frac{\zeta^2}{{\hat T}^2}.$$
(We emphasize that we have slightly changed notation compared to the 
previous subsection. The independent variables for all functions in this 
subsection are ${\hat T}$, ${\hat \nu}$ and $\psi$. The corresponding variables
in the previous subsection were ${\hat T}$, ${\hat \nu}$ and 
$\zeta^2$.)

Note that \eqref{poi} is automatically even under time 
reversal. The corrections induced by \eqref{poi} to the 
stress tensor and {\it consistent} charge current  
(\cite{Bardeen:1984pm}, see section $2.3$ equation $2.16$ of \cite{Banerjee:2012iz}) in equilibrium are given by  
\begin{eqnarray}\label{podstress}
 T^{ij}&=&-\f{2}{\hT} (\zeta^{eq})^{i}(\zeta^{eq})^{j}\left(g_{1,\psi_{eq}} S_1+T_0 g_{2,\psi_{eq}} S_2\right)\nn\\
T_{00}&=& -T_0 e^{\s}\left((-\hT g_{1,\hT}+2 \psi_{eq} g_{1,\psi_{eq}}) S_1+T_0(-\hT  g_{2,\s}+ 2\psi_{eq} g_{2,\psi}) S_2\right)\nn\\
J_{0}&=&- e^{\s}\left(g_{1,\nu} S_1+T_0 g_{2,\nu} S_2\right)-e^{\sigma}\epsilon^{ijk}\bigg[ \frac{C}{3} A_i \nabla_j A_k + \frac{C}{3} A_{0} A_i \nabla_j a_k\bigg]\nn\\
J^{i}&=& \hT \left(2 g_1 (S_1 \f{(\zeta^{eq})^{i}}{\hT^{2} \psi_{eq}}-\f{V_3^{i}}{\hT^{2} \psi_{eq}})+T_0 g_2 (S_2 \f{(\zeta^{eq})^{i}}{\hT^{2} \psi_{eq}}-\f{V_4^{i}}{\hT^{2} \psi_{eq}})+ \hT V_1^{i}g_{1,\hT}-\f{1}{T_0}V_2^{i}g_{1,\nu}-V_{5}^{i}g_{1,\psi_{eq}}\right)\nn\\
&+&\f{2}{\hT}\zeta^{i}(S_1 g_{1,\psi_{eq}}+T_0 S_2 g_{2,\psi_{eq}})\nn\\
&+&e^{-\sigma}\bigg[2 \left(\frac{C}{3} A_{0}\right) \f{1}{\hT^2\psi_{eq}}((\zeta^{eq})^i S_1- V_3^{i}) + \left(\frac{C}{6}A_{0}^{2}\right) \f{1}{\hT^2\psi_{eq}}((\zeta^{eq})^i S_2- V_4^{i}) + \frac{C}{3} \epsilon^{ijk} A_k\nabla_j A_0 \bigg]\nn\\
T_{0}^{i}&=& \hT\Bigg(\f{(T_0 g_2-2 A_0 g_1)}{\hT^{2} \psi_{eq}}(S_1 (\zeta^{eq})^{i}-V_3^{i})-\f{T_0 A_0 g_2}{\hT^{2} \psi_{eq}} (S_2 (\zeta^{eq})^{i}-V_4^{i})+T_0 (\hT V_1^{i}(g_{2,\hT}-\hn g_{1,\hT})\nn\\
&-&\f{1}{T_0}V_{2}^{i}(g_{2,{\hn}}-\hn g_{1,\nu} )-V_{5}^{i}(g_{2,\psi_{eq}}-\hn g_{1,\psi_{eq}}))-\f{2 A_0}{\hT}\zeta^{i}(S_1 g_{1,\psi_{eq}}+T_0 S_2 g_{2,\psi_{eq}})\Bigg)\nn\\
&-& \half C A_{0}^{2}e^{-\sigma}(\f{1}{\hT^2 \psi_{eq}}(\zeta^{eq})^i S_1-\f{1}{\hT^2\psi_{eq}} V_3^{i})+(2 C_1-\frac{C}{6}A_{0}^{3} )e^{-\sigma}(\f{1}{\hT^2\psi_{eq}}(\zeta^{eq})^i S_2-\f{1}{\hT^2\psi_{eq}} V_4^{i})\big],\nn\\
\end{eqnarray}
where 
\begin{eqnarray}\label{oddcur}
\psi_{eq} &=& \frac{\zeta^{eq}_i\zeta^{eq}_j g^{ij}}{\hT^2}\\
 S_1&=&\epsilon^{ijk}\zeta^{eq}_{i}\partial_{j}\zeta^{eq}_{k},~S_2=\epsilon^{ijk}\zeta^{eq}_{i}\partial_{j}a_{k}\nn\\
V_1^{i}&=&\epsilon^{ijk}\zeta^{eq}_{j}\partial_{k}\s,~V_2^{i}=\epsilon^{ijk}\zeta^{eq}_{j}\partial_{k}A_0,~~V_3^{i}=\epsilon^{ijk}\zeta^{eq}_{j}F_{kl}(\zeta^{eq})^{l}\nn\\
V_4^{i}&=&\epsilon^{ijk}\zeta^{eq}_{j}f_{kl}(\zeta^{eq})^{l},~V_5^{i}=\epsilon^{ijk}\zeta^{eq}_{j}\partial_{k}\psi_{eq}\nn\\
V_6^{i}&=&\epsilon^{ijk}F_{jk},~V_7^{i}=\epsilon^{ijk}f_{jk}.\nn\\
\end{eqnarray}
The symbols for $V_{6}^{i}$ and $V_{7}^{i}$ have been introduced for notational 
convenience only; these vectors are determined in terms of the other 
quantities above by 
\begin{eqnarray}
 V_{6}^{i}&=&\f{2}{\hT^2\psi_{eq}}((\zeta^{eq})^i S_1- V_3^{i})\nn\\
V_{7}^{i}&=&\f{2}{\hT^2\psi_{eq}}((\zeta^{eq})^i S_2- V_4^{i}).
\end{eqnarray}
As we have emphasized, the formulas above determine the consistent 
current. The covariant current is obtained from the consistent current
by an additional shift (see section $2.4$ of \cite{Banerjee:2012iz} for a 
review). We find that the one derivative contribution to the 
covariant current in equilibrium is given by 
\begin{equation}\label{covariantcur}\begin{split}
 J_{0}&=- e^{\s}\left(g_{1,{\hn}} S_1+T_0 g_{2,{\hn}} S_2\right)\\
J^{i}&= \hT \left(2 g_1 (S_1 \f{(\zeta^{eq})^{i}}{\hT^{2} \psi_{eq}}-\f{V_3^{i}}{\hT^{2} \psi_{eq}})+T_0 g_2 (S_2 \f{(\zeta^{eq})^{i}}{\hT^{2} \psi_{eq}}-\f{V_4^{i}}{\hT^{2} \psi_{eq}})+ \hT V_1^{i}g_{1,\hT}-\f{1}{T_0}V_2^{i}g_{1,{\hn}}-V_{5}^{i}g_{1,\psi_{eq}}\right)\\
&+\f{2}{\hT}(\zeta^{eq})^{i}(S_1 g_{1,\psi_{eq}}+T_0 S_2 g_{2,\psi_{eq}})\\
&+e^{-\sigma}\bigg[C \f{1}{\hT^2\psi_{eq}}((\zeta^{eq})^i S_1- V_3^{i}) + \left(\frac{C}{2}A_{0}^{2}\right) \f{1}{\hT^2\psi_{eq}}((\zeta^{eq})^i S_2- V_4^{i})  \bigg]\\
\end{split}\end{equation}

\section{Constraints on parity even corrections to constitutive relations at 
first order}

In this subsection we will determine parity even first order corrections to the 
superfluid constitutive relations both from the method of entropy increase 
as well as from the partition function of the previous section, and 
demonstrate their equality. 

Let us first consider the almost trivial case of parity even superfluids 
that also preserve time reversal invariance. As we have explained in the 
previous section, in this case $W_1=0$. It follows immediately from 
this result that all non dissipative superfluid transport coefficients 
must vanish. Exactly this conclusion was reached in \cite{Bhattacharya:2011tra} from the requirement 
of point wise positivity of the divergence of the entropy current in an 
arbitrary fluid flow. 

The study of time reversal non invariant superfluids is more involved.
In this case the constraints from the local second law have not previously 
been analyzed. In this section we first present this analysis. We then 
study the constraints obtained from the analysis of the partition function.
As mentioned above, we will find that these two methods yield identical 
constraints.

\subsection{Constraints from the local second law}

In this subsection (but nowhere else in this paper) we consider the 
non equilibrium flow of a superfluid on a (generically) non stationary 
spacetime. We continue to denote the background metric of our spacetime 
by $G_{\mu \nu}$. The background gauge field is denoted by ${\cal A}_\mu$. The 
variables of superfluid hydrodynamics are the temperature field $T(x^\mu)$, 
velocity field $u^\mu(x^\mu)$ and the gradient of the phase field 
$\xi_\mu= -\partial_\mu \phi + {\cal A}_\mu$. We often work in terms of the 
fluid dynamical field 
$$(\zeta_f)_\mu= \xi_\mu + \mu u_\mu$$
Note that, in equilibrium and at lowest order in the derivative expansion 
$(\zeta_f)_0=0$ and 
$$ (\zeta_f)_i=\xi_i-{\cal A}_0 a_i=\zeta_i.$$

We specify some additional notation that we will use 
extensively below.
\begin{equation}\label{notation2}
\begin{split}
&P^{\mu\nu} = u^\mu u^\nu + G^{\mu\nu},~~\tilde P^{\mu\nu} = P^{\mu\nu} - \frac{(\zeta_f)^\mu(\zeta_f)^\nu}{(\zeta_f)^2},~~V^\mu = \frac{E^\mu}{T} - P^{\mu\nu}\partial_\nu \nu\\
&R = \frac{q}{\epsilon + P},~~~K = \nabla_\mu\left(f \xi^\mu\right) = s (u.\partial)\left(\frac{q}{s}\right),~~~ \Theta =( \nabla.u)=-\frac{u.\partial s}{s}\\
&{\mathfrak a}_\mu = (u.\nabla)u_\mu\\
&H_1 = T,~~ H_2 = \nu, ~~H_3 = (\zeta_f)^2
\end{split}
\end{equation}
In words, $P^{\mu\nu}$ projects onto the three dimensional subspace 
orthogonal to the normal fluid, while $\tilde P^{\mu\nu}$ projects 
onto the two dimensional subspace orthogonal to both the normal and 
superfluid velocities. ${\mathfrak a}_\mu$ and $\Theta$ are the 
normal fluid acceleration and expansion respectively. $V^\mu$ 
is the `Einstein combination' of the electric field and 
derivative of the chemical potential that vanishes in equilibrium. 
$H_1$, $H_2$ and $H_3$ are new names for the three scalar hydrodynamical 
fields; note that $H_2$ is $\nu=\frac{\mu}{T}$ rather than the chemical 
potential itself. Finally $K$ is the term that is set to zero by the 
first order equation of motion of the Goldstone phase, while $R$ is 
a combination of zero order thermodynamical fields that often appears
in the formulas below. 

In order to analyze the constraints that follow from the local form 
of the second law, we follow the procedure described in section $3$ of 
\cite{Bhattacharya:2011tra}. Briefly, we first write down the most general onshell independent
first order entropy current allowed by symmetry. We then compute the 
divergence of this current (this is mere algebra) and then use the 
equations of hydrodynamics, together with the corrected constitutive 
relations
\begin{eqnarray} \label{ltt}
 T^{\mu\nu} &=& (\epsilon+ p)u^{\mu}u^{\nu}+ p G^{\mu\nu}+ f \xi^{\mu}\xi^{\nu}\nn
+ \pi^{\mu\nu}\\
J^{\mu}&=&q u^{\mu}-f \xi^{\mu} + j^\mu,
\end{eqnarray}
(here $\pi^{\mu\nu}$ and $j^\mu$ refer to as yet unspecified one and higher 
derivative corrections to the constitutive relations)
to re express this 
divergence as the sum of a linear form in onshell independent two 
derivative data and a quadratic form in onshell independent one derivative 
data. Point wise positivity of the divergence requires the linear form to 
vanish (this imposes several constraints on the entropy current). 
Once these conditions are imposed, the divergence of the entropy current is 
purely a quadratic form in one derivative data. We require this quadratic form 
to be positive definite. This requirement 
further constrains the entropy current as well as the first order 
contributions to $\pi^{\mu\nu}$ and $j^\mu$ in a manner we now schematically 
describe. 

As we will see below, the quadratic form so obtained has the property that 
it vanishes when projected onto a subset of one derivative terms. In other 
words, all independent one derivative terms can be divided into $y$ type 
`entropically dissipative' terms and $x$ type entropically 
nondissipative terms, and the quadratic form takes the schematic form 
$$A_{i j} y^i y^j + B_{i m} y^i x^m$$
Note that the structure of this quadratic form is preserved under $x$ 
redefinitions 
$$x_m \rightarrow x_m + C_{mi} y^i$$
but not under analogous redefinitions of $y^i$. In other words there exists 
a well defined subspace of dissipative data but no definite subspace 
of nondissipative data.

Positivity of the quadratic form described above requires that 
$A_{ij}$ is a positive matrix, and  $B_{im}=0$ for all $i$ and $m$.
The last set of constraints yield relations between otherwise apparently
independent transport coefficients. 
\footnote{Assuming that the matrix $A$ is positive definite, entropy is 
always produced whenever any of the $y^i$ are nonzero. It follows that all 
$y^i$ must always vanish in equilibrium. This observation motivates 
the following definition, utilized in \cite{Banerjee:2012iz}.
Expressions that vanish in (arbitrary stationary) equilibrium are referred to 
as dissipative data. It follows from that entropically dissipative data
is necessarily dissipative. However the converse is not necessarily true; 
it is possible for data to vanish in arbitrary stationary equilibrium 
but yet be entropically nondissipative. We will see an example of this 
phenomenon later in this paper. }

In order to actually implement this process we need first to choose a basis 
for onshell independent data.  As explained in \cite{Bhattacharya:2011tra} (see 
e.g. Table $3$), at first order in the derivative expansion 
there exist 7 (4 dissipative and 3 non dissipative) onshell independent 
scalars , 7 (2 dissipative and 5 nondissipative) onshell 
independent vectors and 2 (1 dissipative and one nondissipative) 
independent tensors constructed out of thermodynamical fields and background
fields. For the purposes of this section, we will find it useful to choose 
our onshell independent basis as follows.

{\it Basis of independent scalars}:
$$\frac{V.(\zeta_f)}{(\zeta_f)^2},~~(u.\partial H_a), ~~((\zeta_f).\partial H_a),~~~a=\{1,2,3\}$$
The four of these scalars are dissipative (they vanish in equilibrium) 
while the remaining three are nondissipative 
(they are non vanishing in equilibrium, and do not cause entropy production).

{\it Basis of independent vectors}: 

$$\tilde P^{\mu\alpha} V_{\alpha},~~ \tilde P^{\mu\alpha}(\zeta_f)_\beta\sigma^{\beta}_\alpha,~~{\tilde P}_\alpha^\mu (\zeta_f)_\nu f^{\nu \alpha},~~
{\tilde P}_{\alpha \mu} (\zeta_f)_\nu F^{\nu \alpha},~~\tilde P^{\mu\alpha}\partial_\alpha H_a,~~~a=\{1,2,3\}$$ 
The first two vectors are dissipative (they vanish in equilibrium) and 
the remaining five vectors are nondissipative.

{\it Basis of independent tensors}
\begin{equation*}
\begin{split}
&\tilde\sigma_{\mu\nu} = \tilde P_\mu^\alpha\tilde P_\nu^\beta\left[\frac{\nabla_\alpha u_\beta + \nabla_\beta u_\alpha - \tilde P^{\lambda\phi}(\nabla_\lambda u_\phi)\eta_{\alpha\beta}}{2}\right],\\
&\sigma^{(\zeta_f)}_{\mu\nu}= \tilde P_\mu^\alpha\tilde P_\nu^\beta\left[\frac{\nabla_\alpha (\zeta_f)_\beta + \nabla_\beta (\zeta_f)_\alpha - \tilde P^{\lambda\phi}(\nabla_\lambda (\zeta_f)_\phi)\eta_{\alpha\beta}}{2}\right]
\end{split}
\end{equation*}

The first is dissipative (it vanishes in equilibrium) while the second is
nondissipative. 

In this paper we wish to constrain the equations of superfluid hydrodynamics 
presented in a `fluid frame' (see \cite{Bhattacharya:2011eea} for an 
explanation of what this means). Throughout this paper we will further 
restrict our attention to fluid frames with $\mu_{diss}=0$ (again see 
\cite{Bhattacharya:2011eea} for definitions). This choice still permits 
the freedom of field redefinitions of the temperature and normal 
velocity fields (as well as field redefinitions of the superfluid phase, 
as we will exploit later in this paper). Even though we work specifically 
frames in which $\mu_{diss}=0$ our final results may easily be lifted to an 
arbitrary $\mu_{diss}\neq 0$ frame using the frame invariant formalism of 
\cite{Bhattacharya:2011tra}.

The most general form of the entropy current,  consistent with the absence 
of linear two derivative terms in its divergence was determined in \cite{Bhattacharya:2011tra}
(see equation $3.19$ ) and takes the form
\begin{equation} \begin{split} \label{ecf}
J^\mu_S&= J^\mu_{can}+ J^\mu_{new}\\
J^\mu_{can}&=su^\mu - \nu j^\mu - \frac{u_\nu\pi^{\mu\nu}}{T}\\
J^\mu_{new}&=\sum_a c_a (\partial_\nu H_a) Q^{\mu\nu} +  \nabla_\nu (c~ Q^{\mu\nu})\\
&\text{where}~~ Q^{\mu\nu} = f(u^\mu(\zeta_f)^\nu - u^\nu(\zeta_f)^\mu)
\end{split}
\end{equation}

The divergence of $J^\mu_{can}$ was worked out in 
\cite{Bhattacharya:2011eea, Bhattacharya:2011tra} (see for example, equation $3.9$ \cite{Bhattacharya:2011tra}, and recall we work with $\mu_{diss}=0$). 
\begin{equation}\label{ec}
\nabla_\mu J^\mu_{can}=
 -\pi^{\mu\nu}\nabla_\mu\left(\frac{u_\nu}{T}\right) + j^\mu V_\mu +( u_\mu j^\mu)(u.\partial \nu)
\end{equation}
The RHS of \eqref{ec} is given schematically by 
$${(\rm one~derivative~correction~to~constitutive~relation)} \times 
{\rm (entropically dissiaptive~data)},$$ 
\footnote{Note that the one derivative expressions that appear here 
are always entropically dissipative, as 
contributions to changes in the proportional to these one derivative 
expressions yield quadratic terms in entropy production.}
We will now rewrite the RHS of \eqref{ec} as a quadratic form in the basis of 
independent dissipative one derivative data chosen above. 
In order to achieve this we need 
to rewrite all the $y$ type terms  in \eqref{ec} in terms of the 
independent basis of dissipative scalars, vectors and tensors 
listed above. To achieve this we use   the equations of motion
\begin{equation}\label{eom1}
\begin{split}
\frac{(\zeta_f).\partial T}{T} + {\mathfrak a}.(\zeta_f) &= RT (V.(\zeta_f)) - \frac{(\zeta_f)^2 K}{\epsilon +P}\\
\frac{(\zeta_f)_\mu(\zeta_f)_\nu\sigma^{\mu\nu}}{(\zeta_f)^2} + \frac{\Theta}{3} &= -T(1 - \mu R) \frac{V.(\zeta_f)}{(\zeta_f)^2} - \frac{\mu K}{\epsilon +P} - \frac{(u.\partial)(\zeta_f)^2}{2(\zeta_f)^2}
\end{split}
\end{equation}
we find 
\begin{equation}\label{candiv}
\begin{split}
\nabla_\mu J^\mu_{can}
=& -\left(\frac{u_\mu u_\nu \pi^{\mu\nu}}{T^2}\right)(u.\partial T) + (j.u)(u.\partial\nu)
 + \frac{(j.(\zeta_f))(V.(\zeta_f))}{(\zeta_f)^2}\\
&+ \frac{u_\mu(\zeta_f)_\nu \pi^{\mu\nu}}{T}\left[R T \left(\frac{V.(\zeta_f)}{(\zeta_f)^2}\right) - \frac{K}{\epsilon + P}\right] - \frac{1}{2T}\left(\pi^{\mu\nu}\tilde P_{\mu\nu}\right)\Theta\\
&-\frac{1}{T}\left(\pi^{\mu\nu} P_{\mu\nu} - \frac{3}{2}\pi_{\mu\nu}\tilde P^{\mu\nu}\right)\left(\frac{(\zeta_f)_\mu(\zeta_f)_\nu\sigma^{\mu\nu}}{(\zeta_f)^2} + \frac{\Theta}{3}\right)\\
&-2\left(\frac{(\zeta_f)_\alpha \pi^{\alpha\nu}\tilde P_{\nu\mu}\sigma^{\mu\beta}(\zeta_f)_\beta}{T(\zeta_f)^2} \right)+ \left(j^\mu + R u_\alpha\pi^{\alpha\mu}\right)\tilde P_{\mu\nu}V^\nu
-\frac{1}{T}\tilde\sigma_{\mu\nu}\tilde \pi^{\mu\nu}
\\
=&-\left(\frac{u_\mu u_\nu \pi^{\mu\nu}}{T^2}\right)(u.\partial T) + (j.u)(u.\partial\nu)-\left( \frac{\pi^{\mu\nu}\tilde P_{\mu\nu}}{2T}\right)\Theta\\
&+\left(\pi^{\mu\nu}P_{\mu\nu} - \frac{3}{2}\pi^{\mu\nu}\tilde P_{\mu\nu}\right)\left(\frac{u.\partial(\zeta_f)^2}{2T(\zeta_f)^2}\right)\\
&+\left(\frac{V.(\zeta_f)}{(\zeta_f)^2}\right)\left[(j.(\zeta_f)) + R( u_\mu(\zeta_f)_\nu \pi^{\mu\nu}) + (1-\mu R)\left(\pi^{\mu\nu}P_{\mu\nu} - \frac{3}{2}\pi^{\mu\nu}\tilde P_{\mu\nu}\right)\right]\\
&+\left(\frac{K}{\epsilon + P}\right)\left[-\left(\frac{-u_\nu(\zeta_f)_\mu \pi^{\mu\nu}}{T}\right)+ \nu \left(\pi^{\mu\nu}P_{\mu\nu} - \frac{3}{2}\pi^{\mu\nu}\tilde P_{\mu\nu}\right)\right]\\
&-2\left(\frac{(\zeta_f)_\alpha \pi^{\alpha\nu}\tilde P_{\nu\mu}\sigma^{\mu\beta}(\zeta_f)_\beta}{T(\zeta_f)^2} \right)+ \left(j^\mu + R u_\alpha\pi^{\alpha\mu}\right)\tilde P_{\mu\nu}V^\nu-\frac{1}{T}\tilde\sigma_{\mu\nu}\tilde \pi^{\mu\nu}\\
\\
=&\sum_{a=1}^3{\mathfrak S}_a (u.\partial)H_a 
+{\mathfrak S}_4 \left(\frac{V.(\zeta_f)}{(\zeta_f)^2}\right) -2 {\mathfrak V}_2^\nu \left( \frac{\tilde P_{\nu\mu}\sigma^{\mu\beta}(\zeta_f)_\beta}{T(\zeta_f)^2} \right)+ {\mathfrak V}_1^\mu \tilde P_{\mu\nu}V^\nu-\frac{1}{T}\tilde\sigma_{\mu\nu}\tilde \pi^{\mu\nu}\\
\end{split}
\end{equation}
where
\begin{equation}\label{simpscalar}
\begin{split}
{\mathfrak S}_a =& \left[\left(\frac{s}{\epsilon + P}\right)\frac{\partial}{\partial H_a}\left(\frac{q}{s}\right)\right]\left[-\left(\frac{-u_\nu(\zeta_f)_\mu \pi^{\mu\nu}}{T}\right)+ \nu \left(\pi^{\mu\nu}P_{\mu\nu} - \frac{3}{2}\pi^{\mu\nu}\tilde P_{\mu\nu}\right)\right]\\
&+\left(\frac{1}{s}\frac{\partial s}{\partial H_a}\right)\left( \frac{\pi^{\mu\nu}\tilde P_{\mu\nu}}{2T}\right)-\left(\frac{u_\mu u_\nu \pi^{\mu\nu}}{T^2}\right)\delta_{a,1} +(j.u)\delta_{a,2}\\
&+\left(\frac{1}{2 T(\zeta_f)^2}\right)\left(\pi^{\mu\nu}P_{\mu\nu} - \frac{3}{2}\pi^{\mu\nu}\tilde P_{\mu\nu}\right)\delta_{a,3}\\
{\mathfrak S}_4&=\left[(j.(\zeta_f)) + R( u_\mu(\zeta_f)_\nu \pi^{\mu\nu}) + (1-\mu R)\left(\pi^{\mu\nu}P_{\mu\nu} - \frac{3}{2}\pi^{\mu\nu}\tilde P_{\mu\nu}\right)\right]\\
& {\mathfrak V}_2^\nu =(\zeta_f)_\alpha \pi^{\alpha\nu}\\
&{\mathfrak V}_1^\mu =\left(j^\mu + R u_\alpha\pi^{\alpha\mu}\right)\\
&\tilde \pi^{\mu\nu} = \tilde P^{\mu\alpha}\tilde P^{\nu\beta}\left[\pi_{\alpha\beta} -\frac{\eta_{\alpha\beta}}{2}\left(\tilde P_{\theta\phi}\pi^{\theta\phi}\right)\right]\\
&H_1 = T,~~H_2 = \nu , ~~ H_3=  (\zeta_f)^2\\
\end{split}
\end{equation}
The last line of \eqref{candiv} is the final result of this manipulation. 
It expresses the divergence of the entropy current as a linear sum over the 
four dissipative onshell scalars and two dissipative onshell vectors and 
one dissipative tensor listed earlier in this subsection. 
These expressions appear 
multiplied by frame invariant linear combinations of $\pi^{\mu\nu} $ 
and $j^\mu$. 

The frame invariant quantities ${\mathfrak S}_a$ and ${\mathfrak V}_a$
will be used extensively below. For later use we will find it useful to 
regard these quantities as linear functions of $\pi^{\mu\nu}$ and 
$j^\mu$, i.e. 
\begin{equation}\label{linfncal}
{\mathfrak S}_a={\mathfrak S}_a(\pi^{\mu\nu}, j^\mu), ~~~
{\mathfrak V}_a={\mathfrak V}_a(\pi^{\mu \nu}, j^\mu)
\end{equation}

The divergence of the `new' part of the entropy current, $J^\mu_{new}$ 
(see \eqref{ecf})  is 
given by 
\begin{equation}\label{neweven}
\begin{split}
&\nabla_\mu J^\mu_{new}\\
=& \sum_{(a,b)}f\left(\frac{\partial c_a}{\partial H_b} - \frac{\partial c_b}{\partial H_a}\right)((\zeta_f).\partial H_a)(u.\partial H_b) + \sum_a (\partial_\nu H_a)\nabla_\mu Q^{\mu\nu}\\
=& \sum_{(a,b)}f\left(\frac{\partial c_a}{\partial H_b} - \frac{\partial c_b}{\partial H_a}\right)((\zeta_f).\partial H_a)(u.\partial H_b) 
+\sum_a(\partial_\mu H_a )\tilde P^\mu_\nu(\nabla_\alpha Q^{\alpha\nu})\\
&- \sum_a\left[(u.\partial H_a) \left(u_\nu\nabla_\mu Q^{\mu\nu}\right) +((\zeta_f).\partial H_a)\left(\frac{ (\zeta_f)_\nu\nabla_\mu Q^{\mu\nu}}{(\zeta_f)^2}\right)\right]
\end{split}
\end{equation}
where $Q^{\mu\nu}$ was defined in \eqref{ecf}. 

Using  equations of motion we can express $\left(u_\nu\nabla_\mu Q^{\mu\nu}\right)$,  $\left(\frac{ (\zeta_f)_\nu\nabla_\mu Q^{\mu\nu}}{(\zeta_f)^2}\right)$ and $\tilde P^\mu_\nu(\nabla_\alpha Q^{\alpha\nu})$ in terms of the onshell independent 
basis scalars of this subsection
 ( spanned by $(u.\partial H_a),~~((\zeta_f).\partial H_a),~~
\left(\frac{V.(\zeta_f)}{(\zeta_f)^2}\right)$)  and vectors (spanned by  $\tilde P^{\mu\alpha} V_{\alpha},~~ \tilde P^{\mu\alpha}(\zeta_f)_\beta\sigma^{\beta}_\alpha,~~\tilde P^{\mu\alpha}\partial_\alpha H_a$). 

\begin{equation}\label{eom3}
\begin{split}
\left(u_\nu\nabla_\mu Q^{\mu\nu}\right)&=s(u.\partial)\left(\frac{f\mu}{s}\right)+\left(1-\frac{f(\zeta_f)^2}{\epsilon + P}\right)K + f\left(\frac{(\zeta_f).\partial T}{T}\right) - f (V.(\zeta_f))\\
\left(\frac{ (\zeta_f)_\nu\nabla_\mu Q^{\mu\nu}}{(\zeta_f)^2}\right)&=s(u.\partial)\left(\frac{f}{s}\right) + fT\left[(1 - \mu R)\left(\frac{V.(\zeta_f)}{(\zeta_f)^2}\right) + \frac{\nu K}{\epsilon +P} + \frac{u.\partial(\zeta_f)^2}{T(\zeta_f)^2}\right]\\
\tilde P^\mu_\nu(\nabla_\alpha Q^{\alpha\nu})&=-P^\mu_\nu\sum_a f c_a\left[T(1-\mu R)V^\nu + 2 (\zeta_f)_\alpha\sigma^{\nu\alpha}\right]
\end{split}
\end{equation}
From equations \eqref{candiv}, \eqref{simpscalar}, \eqref{neweven} and \eqref{eom3} we conclude that, no matter what form the fluid constitutive relations 
take, the divergence of the entropy current cannot contain any expressions 
 of the form $((\zeta_f).\partial H_a)^2$ or $(\tilde P^{\mu\nu}\partial_\mu H_a \partial_\nu H_b)$. In other words the scalars $(\zeta_f). \partial H_a$ and the 
vectors $(\tilde P^{\mu\nu}\partial_\mu H_a)$ are nondissipative. 
It follows that the positivity of $(\nabla_\mu J^\mu_S)$ requires that 
the divergence contain no term linear in 
$((\zeta_f).\partial H_a)$ or $(\tilde P^{\mu\nu}\partial_\mu H_a)$ (see e.g. \cite{Bhattacharya:2011tra} for repeated use of similar arguments.) 
To ensure this $\pi^{\mu\nu}$ and $j^\mu$ have to satisfy the following conditions.
\begin{equation}\label{evencond}
\begin{split}
{\mathfrak S}_a =& ~-\sum_{b=1}^3 ((\zeta_f).\partial H_b)\bigg\{f\left(\frac{\partial c_b}{\partial H_a} - \frac{\partial c_a}{\partial H_b}\right)-\frac{f c_a}{T} \delta_{b,1} + \frac{f c_b}{(\zeta_f)^2}\delta_{a,3}\\
& ~~~~~~~~~~~~~~+c_b \left[s \frac{\partial}{\partial H_a}\left(\frac{f}{s}\right)+\left( \frac{s\nu}{\epsilon +P}\right) \frac{\partial}{\partial H_a}\left(\frac{q}{s}\right) \right]\bigg\}\\
&+ \text{dissipative terms}\\
=&~-\sum_{b=1}^3 ((\zeta_f).\partial H_b)\bigg\{\left[\frac{\partial}{\partial H_a}(f c_b)
 - \frac{f}{T}\frac{\partial}{\partial H_b}(T c_a)\right]+ \frac{f c_b}{(\zeta_f)^2}\delta_{a,3}\\
& ~~~~~~~~~~~~~~+c_b \left[-\frac{1}{s} \frac{\partial s}{\partial H_a}+\left( \frac{s\nu}{\epsilon +P}\right) \frac{\partial}{\partial H_a}\left(\frac{q}{s}\right) \right] \bigg\}\\
&+ \sum_{b=1}^3 M_{ab} (u.\partial H_b) + M_{a4}\left(\frac{V.(\zeta_f)}{(\zeta_f)^2}\right)~~~~~~~~~~(a=\{1,2,3\})\\
{\mathfrak S}_4 =& (j.(\zeta_f)) + R( u_\mu(\zeta_f)_\nu \pi^{\mu\nu}) + (1-\mu R)\left(\pi^{\mu\nu}P_{\mu\nu} - \frac{3}{2}\pi^{\mu\nu}\tilde P_{\mu\nu}\right) + \text{dissipative terms}\\
=&-\sum_b((\zeta_f).\partial H_b)  f T (1-\mu R)c_b +\sum_{b=1}^3 M_{4b}(u.\partial H_b) + M_{44}\left(\frac{V.(\zeta_f)}{(\zeta_f)^2}\right)\\
\end{split}
\end{equation}
\begin{equation}\label{evencond1}
\begin{split}
{\mathfrak V}_{1\mu}=&\left(j^\nu + R u_\alpha\pi^{\alpha\nu}\right)\tilde P_{\mu\nu}= T(1-\mu R)f\sum_b  c_b (\tilde P_\mu^\nu\partial_\nu H_b)+ \text{dissipative terms}\\
=&~T(1-\mu R)f\sum_b  c_b (\tilde P_\mu^\nu\partial_\nu H_b)+ N_{11} \left(\tilde P_{\mu\nu} V^\nu\right) -N_{12}\left(\frac{\tilde P_{\mu\beta}(\zeta_f)_\alpha\sigma^{\alpha\beta}}{2T(\zeta_f)^2}\right)\\
\\
{\mathfrak V}_{2\mu}=&(\zeta_f)_\alpha \pi^{\alpha\nu}\tilde P_{\nu\mu}=-T(\zeta_f)^2f\sum_b  c_b (\tilde P_\mu^\nu\partial_\nu H_b)+ \text{dissipative terms}\\
=&~-T(\zeta_f)^2f\sum_b  c_b (\tilde P_\mu^\nu\partial_\nu H_b)+ N_{21} \left(\tilde P_{\mu\nu} V^\nu\right) -N_{22}\left(\frac{\tilde P_{\mu\beta}(\zeta_f)_\alpha\sigma^{\alpha\beta}}{2T(\zeta_f)^2}\right)\\
\tilde \pi^{\mu\nu} = &~\tilde P^{\mu\alpha}\tilde P^{\nu\beta}\left[\pi_{\alpha\beta} -\frac{\eta_{\alpha\beta}}{2}\left(\tilde P_{\theta\phi}\pi^{\theta\phi}\right)\right] = \text{dissipative term}=-\eta~ \tilde \sigma^{\mu\nu}\\
\end{split}
\end{equation}
where $M$ is a $4\times 4$ matrix of dissipative transport coefficients in the scalar sector and $N$ is a  $2\times 2$ matrix of dissipative transport coefficients in the vector sector.

Equations \eqref{evencond},\eqref{evencond1} are the main result of this subsection. It expresses the 
equality type constraints that follow from the local second law. 
Once \eqref{evencond},\eqref{evencond1} are satisfied the final expressions for the divergence of the entropy current takes the following form.
\begin{equation}\label{finaldiv}
\begin{split}
&\nabla_\mu J^\mu_s \\
=&~ -\sum_{a,b} c_a\left\{\mu s\frac{\partial \left(\frac{f}{s}\right) }{\partial H_b}+ s\left(1 - \frac{f(\zeta_f)^2}{\epsilon +p}\right)\frac{\partial \left(\frac{q}{s}\right)}{\partial H_b} + f\left( T \delta _{b,2} + \nu \delta_{b,1}\right)\right\}(u.\partial H_a)(u.\partial H_b)\\
&-f(V.(\zeta_f))\sum_a c_a (u.\partial H_a)\\
&+\sum_{a,b=1}^3 M_{ab} (u.\partial H_a)(u.\partial H_b) + \sum_{a=1}^3 \left(M_{4a}+M_{a4}\right)(u.\partial H_a)\left(\frac{V.(\zeta_f)}{(\zeta_f)^2}\right) + M_{44} \left(\frac{V.(\zeta_f)}{(\zeta_f)^2}\right)^2\\
&+\tilde P_{\mu\nu}\left[N_{11} \left(V^\mu V^\nu  \right) + (N_{12} + N_{21}) (V^\mu (\zeta_f)_\alpha\sigma^{\alpha\nu}) + N_{22} ( (\zeta_f)_\alpha(\zeta_f)_\beta\sigma^{\alpha\mu}\sigma^{\beta\nu})\right]\\ &+\frac{\eta}{T}\tilde\sigma_{\mu\nu}\tilde\sigma^{\mu\nu}
\end{split}
\end{equation}
The positivity of this quadratic form imposes additional inequality 
type constraints on transport coefficients that we will not further 
explore here. 

\subsection{Constraints from the partition function}

In this subsection we now reproduce the conditions \eqref{evencond},\eqref{evencond1} using 
considerations independent of those of the previous subsection. 
 The procedure we adopt is very similar to that described in 
\cite{Banerjee:2012iz}, and we describe it only briefly, highlighting only those elements 
of the analysis that are unique to the superfluid. 

The starting point of our analysis is the expressions \eqref{evencurrent} and 
\eqref{evenstress} which represent the first order for the corrections to the  
stress tensor and charge current that follow by varying the 
local action for the Goldstone mode w.r.t. the metric and background 
gauge field. Once we substitute in the solution for the field $\xi^\mu(x)$, 
according to its equations of motion, \eqref{evencurrent} and \eqref{evenstress}
yield first order corrections $\delta T_{\mu\nu}$ and $\delta J^\mu$ 
to the values of the stress tensor and charge current in thermal equilibrium.

From the hydrodynamical point of view, $\delta T_{\mu\nu}$ and 
$\delta J^\mu$ are the first order contributions in  \eqref{ltt} 
once we substitute 
\begin{equation}\label{subeq}
T(x)= {\hat T}(x) + T_1(x), ~~~~\mu(x)={\hat \mu}(x) + \mu_1(x), ~~~
u^\mu(x)= {\hat u}^\mu + u_1^\mu(x)
\end{equation}
into those expressions. Here $T_1(x)$, $\mu_1(x)$ and $u^\mu_1(x)$ a
are the first derivative corrections to the equilibrium configurations of 
temperature, chemical potential and velocity. 

Upon substituting \eqref{subeq} into \eqref{ltt} we get first derivative
contributions of two sorts.  
First we have the corrections to constitutive relations evaluated on 
the zero order equilibrium configurations 
$\Pi^{\mu\nu} ({\hat T}, {\hat \mu}, {\hat u}^\mu, ({\zeta}^{eq})^\mu)$ 
and $j^\mu({\hat T}, {\hat \mu}, {\hat u}^\mu, (\zeta^{eq})^\mu)$ .
Second we have contributions from terms proportional to $T_1$, $\mu_1$ and 
$u^\mu_1$ when \eqref{subeq} is plugged into the perfect fluid constitutive 
relations.
Contributions of the second 
sort, however,  {\it precisely} cancel out in the frame invariant 
linear combinations ${\mathfrak S}_a$ ($a=1 \ldots 4$) and 
${\mathfrak V}_a$ ($a=1 \ldots 2$). In other words 
\begin{equation}\label{logcon} \begin{split}
&{\mathfrak S}_a(\delta T_{\mu\nu}, \delta J_\mu)
={\mathfrak S}_a(\pi_{\mu\nu}, j_\mu)\\
&{\mathfrak V}_a(\delta T_{\mu\nu}, \delta J_\mu)
={\mathfrak V}_a(\pi_{\mu\nu}, j_\mu)\\
\end{split}
\end{equation}
($\pi^{\mu\nu}$ and $j^\mu$ on the RHS of \eqref{logcon} are evaluated 
on the zero order equilibrium configurations). In the general formulation 
of hydrodynamics, however, it is precisely the frame invariants 
that appear on the RHS of \eqref{logcon} that are 
expanded in the most general symmetry allowed 
constitutive relations (see e.g. \cite{Bhattacharya:2011tra} )
\begin{equation}\label{focc} \begin{split}
{\mathfrak S}_a (\pi^{\mu\nu}, j^\mu)&={\alpha}_{am} S^m ~~~(a=1 \ldots 4), ~~~(m=1 \ldots 7)\\
{\mathfrak V}_a^\mu(\pi^{\mu\nu}, j^\mu)&={\gamma}_{am} V_m^\mu~~~(a=1 \ldots 2), ~~~(m=1 \ldots 5)\\
\end{split}
\end{equation}
where $S^m$ and $V_m^\mu$ are the independent one derivative scalars and 
vectors and the coefficients $\alpha_{a m}$ and $\gamma_{a m}$ are arbitrary 
functions of the scalars ${T}$, ${\mu}$ and 
${\xi}^\mu {\xi}_\mu$.  

$\alpha_{am}$ and $\gamma_{a m}$ are the 
constitutive coefficients we wish to constrain, and this is achieved as 
follows. In the LHS of \eqref{logcon} we substitute the expressions 
 \eqref{evencurrent} and \eqref{evenstress}) for $\delta T^{\mu\nu}$ and $\delta J^\mu$. This determines the LHS of 
\eqref{logcon} completely in terms of the functions $f_1$, $f_2$ and $f_3$ that 
appear in the partition function. In the RHS of \eqref{logcon} 
we substitute  \eqref{focc}, and evaluate these expressions in equilibrium 
$$T={\hat T}, ~~ \mu={\hat \mu}, ~~ \zeta=\zeta^{eq}.$$
Under the last substitution those of $S^m$ and $V^m$ that are dissipative 
vanish. The non dissipative one derivative scalars and vectors evaluate 
to geometric expressions. Equating the coefficients of these expressions 
we determine $\alpha_{am}$ and $\gamma_{a m}$ for those values of $m$ 
that correspond to non dissipative terms. In other words this 
procedure completely determines all non dissipative transport coefficients. 
\footnote{There is an important subtlety here. All of the operations described 
above
may only be performed in equilibrium, i.e. once we have solved for
$(\zeta^{eq})^\mu$ as a function of background fields and substituted this back 
into the partition function. We implement our programme without 
explicitly solving, simply by treating 
$(\zeta^{eq})^\mu(x)$ as formally independent of the other background fields, 
except for those local combinations of $(\zeta^{eq})^\mu$ that appear in terms 
of its equation of motion and derivatives there off. The reason for this that 
the expressions for $\xi^\mu$ as a function of background fields is highly 
nonlocal. The 
only situation in which cancellations are possible between local expressions 
in $(\zeta^{eq})^\mu$ and local expressions in the background fields is when 
we get derivatives combining with $({\zeta}^{eq})^\mu$ in the form of the $\phi$ equations of 
motion.}

In the rest of this subsection we implement the procedure described above 
to explicitly determine all nondissipative transport coefficients in terms 
of the three free functions $f_1$, $f_2$ and $f_3$ that enter the 
local action for the Goldstone field. We demonstrate that our results agree 
exactly with \eqref{evencond},\eqref{evencond1}, obtained from the local form of the second
law, once we identify the three unknown functions $c_1$, $c_2$ and $c_3$ 
in the entropy current of the previous subsection in terms of the functions in 
the partition function according to 
\begin{equation}\label{idparameter}
c_1 = \frac{f_1}{fT} + \frac{1}{T}\frac{\partial f_3}{\partial T},
~~~c_2 = \frac{f_2}{fT} + \frac{1}{T}\frac{\partial f_3}{\partial \nu},~~~
c_3 =  \frac{1}{T}\frac{\partial f_3}{\partial \zeta^2}
\end{equation}
We will also demonstrate that the identification \eqref{idparameter}
may be argued for directly by comparing the thermodynamical 
 entropy in equilibrium with 
the integral of the equilibrium entropy current over a spatial slice.

It will be useful in the computation below to note that 
$P^{\mu\nu}$ and $\tilde P^{\mu\nu}$ are given by                           
 $$P_{ij} = g_{ij},~~~\tilde P_{ij} =g_{ij} - \frac{\zeta^{eq}_i\zeta^{eq}_j}{(\zeta^{eq})^2}$$

We turn now to the explicit computation, starting with the vectors.
\begin{equation}\label{deriv1}
\begin{split}
{\mathfrak V}_{10}(\delta T_{\mu\nu}, \delta J_\mu) =&~ 
{\mathfrak V}_{20}(\delta T_{\mu\nu}, \delta J_\mu) =0\\
{\mathfrak V}_{1i}(\delta T_{\mu\nu}, \delta J_\mu)=&~\tilde P_{ij}\left(\delta J^j + {\hat R} {\hat u}^0 \delta T^j_0\right)\\
=&~ \tilde P_{ij}\left(\delta J^j - R e^{-\sigma}A_0 \delta J^j\right)\\
=&~(1-\hm {\hat R})\tilde P_{ij}\delta J^j\\
=&~\tilde P_{ij}g^{jk}(1-\hm {\hat R})\left(\frac{f_1}{\hT}\partial_k \hT +\frac{f_2}{\hT}\partial_k \hn +\frac{f}{\hT}\partial_k f_3\right)\\
{\mathfrak V}_{2i}(\delta T_{\mu\nu}, \delta J_\mu) =&~\tilde P_{ij}\zeta_k ~\delta T^{kj}= -\zeta^2 \tilde P_{ij}\delta J^j\\
=&~-\tilde P_{ij}g^{jk}\left(\frac{f_1}{\hT}\partial_k \hT +\frac{f_2}{\hT}\partial_k \hn +\frac{f}{\hT}\partial_k f_3\right)\\
\end{split}
\end{equation}
The last line of \eqref{deriv1} exactly matches \eqref{evencond},\eqref{evencond1}) upon 
using the identification of the parameters \eqref{idparameter}.

We turn next to the scalars; let us start with ${\mathfrak S}_4$.
\begin{equation}\label{deri4}
\begin{split}
{\mathfrak S}_4(\delta T_{\mu\nu}, \delta J_\mu) =&~(j.\zeta) + R( u_\mu\zeta_\nu \pi^{\mu\nu}) + (1-\mu R)\left(\pi^{\mu\nu}P_{\mu\nu} - \frac{3}{2}\pi^{\mu\nu}\tilde P_{\mu\nu}\right)\\
=&~-(1-\hm {\hat R} ) \left[ f_1(\zeta^{eq}.\partial \hT) + f_2(\zeta^{eq}.\partial \hn) + f(\zeta^{eq}.\partial f_3) \right]\\
=&~\hT(1-\hm {\hat R})\sum_b f c_b (\zeta^{eq}.\partial H_b)
\end{split}
\end{equation}
In the last step we have used \eqref{idparameter}, and have obtained manifest 
agreement with \eqref{evencond},\eqref{evencond1}.

Next we shall calculate the remaining three scalars ${\mathfrak S}_a,~~~~a=\{1,2,3\}$. The algebraic manipulations here are a little more involved than in 
previous cases, and we provide some details. 
\begin{equation}\label{deriv2}
\begin{split}
{\mathfrak S}_a(\delta T_{\mu\nu}, \delta J_\mu) =& \left[\left(\frac{s}{\epsilon + P}\right)\frac{\partial}{\partial H_a}\left(\frac{q}{s}\right)\right]\left[-\left(\frac{-u_\nu\zeta_\mu \pi^{\mu\nu}}{T}\right)+ \nu \left(\pi^{\mu\nu}P_{\mu\nu} - \frac{3}{2}\pi^{\mu\nu}\tilde P_{\mu\nu}\right)\right]\\
&+\left(\frac{1}{s}\frac{\partial s}{\partial H_a}\right)\left( \frac{\pi^{\mu\nu}\tilde P_{\mu\nu}}{2T}\right)-\left(\frac{u_\mu u_\nu \pi^{\mu\nu}}{T^2}\right)\delta_{a,1} +(j.u)\delta_{a,2}\\
&+\left(\frac{1}{2 T\zeta^2}\right)\left(\pi^{\mu\nu}P_{\mu\nu} - \frac{3}{2}\pi^{\mu\nu}\tilde P_{\mu\nu}\right)\delta_{a,3},~~~~~( a= \{1,2,3\})\\
\end{split}
\end{equation}
The first line in \eqref{deriv2} can be evaluated as
\begin{equation}\label{step1}
\begin{split}
&\left[\left(\frac{s}{\epsilon + P}\right)\frac{\partial}{\partial H_a}\left(\frac{q}{s}\right)\right]\left[-\left(\frac{-u_\nu\zeta_\mu \pi^{\mu\nu}}{T}\right)+ \nu \left(\pi^{\mu\nu}P_{\mu\nu} - \frac{3}{2}\pi^{\mu\nu}\tilde P_{\mu\nu}\right)\right]\\
=&~\left[\left(\frac{s}{\epsilon + P}\right)\frac{\partial}{\partial H_a}\left(\frac{q}{s}\right)\right]
\left[-\hn f (\zeta^{eq}.\partial f_3)-\hn f_2 (\zeta^{eq}.\partial \hn)-\hn f_1 (\zeta^{eq}.\partial \hT)\right]\\
=&~-\left[\left(\frac{s\hn}{\epsilon + P}\right)\frac{\partial}{\partial H_a}\left(\frac{q}{s}\right)\right]\sum_b c_b (\zeta^{eq}.\partial H_b)
\end{split}
\end{equation}
In the last step we have used \eqref{idparameter}.

The second line of \eqref{deriv2} may be evaluated as follows
\begin{equation}\label{step2}
\begin{split}
&\left(\frac{1}{s}\frac{\partial s}{\partial H_a}\right)\left( \frac{\pi^{\mu\nu}\tilde P_{\mu\nu}}{2T}\right)\\
=~&\left(\frac{1}{s}\frac{\partial s}{\partial H_a}\right)\left[\frac{f_1}{\hT}(\zeta^{eq}.\partial)\hT + \frac{f_2}{\hT}(\zeta^{eq}.\partial)\hn + \frac{f}{\hT}(\zeta^{eq}.\partial)f_3\right] \\
=~&\left(\frac{1}{s}\frac{\partial s}{\partial H_a}\right)\sum_b f c_b (\zeta^{eq}.\partial H_b)
\end{split}
\end{equation}
In the last step we have used \eqref{idparameter}.

Finally we evaluate the last three terms of \eqref{deriv2} together.
\begin{equation}\label{step3}
\begin{split}
&-\left(\frac{u_\mu u_\nu \pi^{\mu\nu}}{T^2}\right)\delta_{a,1} +(j.u)\delta_{a,2}
+\left(\frac{\pi^{\mu\nu}P_{\mu\nu} - \frac{3}{2}\pi^{\mu\nu}\tilde P_{\mu\nu}}{2 T\zeta^2}\right)\delta_{a,3}\\
=~&-\left[\frac{\partial}{\partial \hT}\left(\frac{f_1}{\hT}\right)(\zeta^{eq}.\partial \hT) + \frac{\partial}{\partial \hT}\left(\frac{f_2}{\hT}\right)(\zeta^{eq}.\partial\nu)+\frac{\partial}{\partial \hT}\left(\frac{f}{\hT}\right)(\zeta^{eq}.\partial f_3)-\frac{f}{\hT}\zeta^{eq}.\partial\left(\frac{f_1}{f}\right)\right]\delta_{a,1}\\
&-\left[\frac{\partial}{\partial \nu}\left(\frac{f_1}{\hT}\right)(\zeta^{eq}.\partial \hT) + \frac{\partial}{\partial \nu}\left(\frac{f_2}{\hT}\right)(\zeta^{eq}.\partial\nu)+\frac{\partial}{\partial \nu}\left(\frac{f}{\hT}\right)(\zeta^{eq}.\partial f_3)-\frac{f}{\hT}\zeta^{eq}.\partial\left(\frac{f_2}{f}\right)\right]\delta_{a,2}\\
&-\frac{1}{\hT}\left[\frac{\partial}{\partial(\zeta^{eq})^2 }\left(\frac{f_1}{\hT}\right)(\zeta^{eq}.\partial \hT) + \frac{\partial}{\partial (\zeta^{eq})^2}\left(\frac{f_2}{\hT}\right)(\zeta^{eq}.\partial\nu)+\frac{\partial}{\partial (\zeta^{eq})^2}\left(\frac{f}{\hT}\right)(\zeta^{eq}.\partial f_3)\right]\delta_{a,3}\\
&-\left[\frac{f_1}{\hT}(\zeta^{eq}.\partial \hT) + \frac{f_2}{\hT}(\zeta^{eq}.\partial \nu)+ \frac{f}{\hT}(\zeta^{eq}.\partial f_3)\right]\delta_{a,3}\\
\\
=~&\sum_b(\zeta^{eq}.\partial H_b)\left[-\left(\frac{\partial f_3}{\partial H_b}\right)\frac{\partial}{\partial H_a}\left(\frac{f}{\hT}\right) +\frac{f}{\hT}\frac{\partial}{\partial H_b}\left(\frac{f_1}{f}\right)\delta_{a,1} +\frac{f}{\hT}\frac{\partial}{\partial H_b}\left(\frac{f_2}{f}\right)\delta_{a,2}\right]\\
&-\frac{\partial}{\partial H_a}\left(\frac{f_1}{\hT}\right)(\zeta^{eq}.\partial \hT)-\frac{\partial}{\partial H_a}\left(\frac{f_2}{\hT}\right)(\zeta^{eq}.\partial \nu) \\ 
&-\left[\frac{f_1}{\hT}(\zeta^{eq}.\partial \hT) + \frac{f_2}{\hT}(\zeta^{eq}.\partial \nu)+ \frac{f}{\hT}(\zeta^{eq}.\partial f_3)\right]\delta_{a,3}\\
\\
=~&-\sum_b(\zeta^{eq}.\partial H_b)\left[\frac{\partial}{\partial H_a}(f c_b) -\frac{f}{\hT}\frac{\partial}{\partial H_b}(\hT c_a) +\frac{ f c_b}{(\zeta^{eq})^2}~ \delta _{a,3} \right]
\end{split}
\end{equation}
In the last step we have used \eqref{idparameter}.

Combining \eqref{step1}, \eqref{step2} and \eqref{step3} 
it is straightforward to verify that the expressions for 
${\mathfrak S}_a,~~~a=\{1,2,3,4\}$ as derived from partition function in this 
subsection, match exactly with \eqref{evencond},\eqref{evencond1}. Note that both methods 
leave dissipative contributions to constitutive relations completely 
unconstrained. 

\subsection{Entropy from the partition function}

In this subsection we will explain how the nondissipative part of the 
entropy current of the superfluid may be read off in a rather direct 
way from the partition function. Our analysis is largely structural, 
and applies equally well to normal (non super) fluids. However our 
presentation applies only at first order in the derivative expansion. 

For any system the entropy $S_T$ in equilibrium may be evaluated from the 
logarithm of partition function $W= \ln Z$ via the thermodynamical relation
\begin{equation}\label{evenentropy}
\begin{split}
S_T &=  W+ T_0\frac{ \partial { W}}{\partial T_0}\\
\end{split}
\end{equation}
We will now rewrite this expression in terms of the goldstone action that 
generates the partition function. Let this action take the form 
$$ S= \int \sqrt{g} {\cal L} d^3 x $$
and also suppose $${\cal L}^{eq} = {\cal L}(\zeta_\mu = \zeta_\mu^{eq})$$
Now we can think of the partition function as
$$W=S({\hat T}, {\hat \mu},T_0 a_i, \zeta^{eq}_\mu)=\int \sqrt{g} ~{\cal L}^{eq}({\hat T}, {\hat \mu},T_0 a_i, \zeta^{eq}_\mu) d^3 x  $$
Using the simple rescaling of the time coordinate employed in 
subsection 2.3.1 of \cite{Banerjee:2012iz} one may show that 
\begin{equation} \begin{split}
T_0 \frac{\partial{\hat T}}{\partial T_0}&= {\hat T}\\ 
T_0 \frac{\partial{\hat \nu}}{\partial T_0}&= -{\hat \nu}\\
T_0 \frac{\partial{a_i}}{\partial T_0}&= 0\\
T_0 \frac{\partial{\zeta_i}}{\partial T_0}&= 0\\
\end{split}
\end{equation}
It follows that 
\begin{equation}\label{somestep}
\begin{split}
&\frac{ \partial {W}}{\partial T_0}\\
=&\int d^3y \sqrt{g} \left[\left(\frac{\delta {\cal L}^{eq}}{\delta\hT(y)}\right)\left(\frac{\partial\hT(y)}{\partial T_0}\right) + \left(\frac{\delta {\cal L}^{eq}}{\delta\hn(y)}\right)\left(\frac{\partial\hn(y)}{\partial T_0}\right) 
+\left(\frac{\delta {\cal L}^{eq}}{\delta a_i(y)}\right)
\left(\frac{a_i(y)}{ T_0}\right)\right]\\
=&\int d^3y \sqrt g e^{-\sigma} \left[\frac{T_{00}}{T_0^2} + 
\frac{\hn  J_0}{T_0} + a_i \left(\frac{T^i_0 + A_0 \delta J^i}{\hT^2}\right)
\right]\\
=&\int d^3y\sqrt g e^\sigma \left[\frac{1}{T_0^2}\left(T_{00} e^{-2\sigma}+ a_i T^i_0\right) +\frac{\hn}{T_0}\left(J_0 e^{-2\sigma} + a_i J^i\right)\right]\\
=&\int d^3y \sqrt {-G}\left[\frac{1}{T_0^2}\left(-\frac{T_{00}}{G_{00}} + 
\frac{G_{0i}}{G_{00}} T^i_0\right) + \frac{\hn}{T_0}\left(-\frac{J_{0}}{G_{00}} + \frac{G_{0i}}{G_{00}} J^i\right)\right]\\
=&\int d^3y\frac{\sqrt {- G}}{T_0}\left[ -\frac{T_0 ^0}{T_0} -\hn J^0\right]\\
=&\int d^3y \frac{\sqrt{-G}}{T_0} \left [- \frac{\hat u^\mu T_\mu ^0}{\hT}-
\hn J^0 \right]
\end{split}
\end{equation}
so that 
\begin{equation}\label{finstep}
S_T= W+\frac{ \partial {W}}{\partial T_0}
= \int d^3y \frac{\sqrt {-G}}{T_0} \left [ {\hat T} {\cal L}^{eq} 
- \frac{\hat u^\mu T_\mu ^0}{\hT}-
\hn J^0 \right]
\end{equation}
This expression may be expanded to first order in derivatives employing
\begin{equation} \begin{split}
{\cal L}^{eq} & = \frac{{\hat P}}{\hat T} +{\cal L}^{eq}_1\\
T_0^0 &=(T^0_0)_{perf} 
+ {\delta T_0^0}\\
J^0&= J^0_{perf} + \delta J^0\\ 
\end{split}
\end{equation} 
where, from \eqref{cc}
\begin{equation} \begin{split}
(T^0_0)_{perf}&=-{\hat \epsilon}  -{\hat f} e^{-2 \sigma} A_0^2 -{\hat f} A_0 a^i 
\zeta_i^{eq} \\
J^0_{perf}&=e^{-\sigma} {\hat q} -{\hat f} (e^{-2 \sigma} A_0 + a^i \zeta_i^{eq}),  
\end{split}
\end{equation} 
$\delta T_0^0$ is defined in \eqref{evenstress}
and $\delta J^0$ is defined in \eqref{evencurrent}. 

Using the Gibbs Duham relation 
$$s= \frac{P+ \epsilon -q \mu}{T}$$
we find that 
\begin{equation}\label{fsn} \begin{split}
S_T&= \int d^3y \sqrt{g} {\hat s} \\ 
&+\int d^3y \frac{\sqrt {-G}}{T_0} \left [ {\hat T} {\cal L}_1^{eq} 
- \frac{\hat u^\mu  \delta T_\mu ^0}{\hT}-
\hn  \delta J^0 \right]\\
\end{split}
\end{equation}
(all proportional to ${\hat f}$ cancel out at zero order in the derivative
expansion). 

Now let us recall that 
$$\delta T^{\mu}_\nu= \left( \pi^\mu_\nu \right)_0+ (T^\mu_\nu)_{perf}^1$$
where $( \pi^\mu_\nu )_0 $ refers to $\pi^\mu_\nu$ evaluated on the zero 
order equilibrium solution and $(T^\mu_\nu)_{perf}^1$ refers to the one derivative
correction in $T^\mu_\nu$ from the first order correction to the equilibrium 
solution. Similarly 
$$\delta J^\mu= \left( j^\mu \right)_0+ (J^\mu_{perf})^1.$$ 
It follows that 
\begin{equation}\label{fsn} \begin{split}
S_T&= \int d^3y  \sqrt{-G} \left[ {\hat s}{\hat u}^0 
-\frac{\hat u^\mu  \delta (T_\mu ^0)_{perf}^1}
{\hT} -
\hn  \delta (J^0)_{perf}^1 \right]  \\ 
&+\int d^3y \frac{\sqrt {-G}}{T_0} \left [ {\hat T} {\cal L}_1^{eq} 
- \frac{\hat u^\mu  \delta \pi_\mu ^0}{\hT}-
\hn  \delta j^0 \right]\\
\end{split}
\end{equation}
so that 
\begin{equation}\label{fsnn} \begin{split}
S_T&= \int d^3y \sqrt{-G} s u^0\\ 
&+\int d^3y \frac{\sqrt {-G}}{T_0} \left [ {\hat T} {\cal L}_1^{eq} 
- \frac{\hat u^\mu  \delta \pi_\mu ^0}{\hT}-
\hn  \delta j^0 \right]\\
\end{split}
\end{equation}
where $su^0$ in \eqref{fsnn} refers to the entropy evaluated on the 
first order corrected solution. In going from \eqref{fsn} to \eqref{fsnn}
we have used the fact that the frame invariance
(see \cite{Bhattacharya:2011tra} for a definition and extensive 
discussion of frame invariance)  of the canonical entropy 
current 
$$J^\mu_{can}=su^\mu - \nu j^\mu - \frac{u_\nu\pi^{\mu\nu}}{T}$$
implies that 
$$ s u^\mu -{\hat s} {\hat u}^\mu +\nu (J^\nu)_{perf}^1+  
\frac{u_\nu (T^{\mu\nu})_{perf}^1}{T}=0.$$
It follows from \eqref{fsnn} that 
\begin{equation}\label{fsm} 
S_T= \int d^3y \sqrt{-G}  \left[ J^0_{can} 
+ \frac{1}{T_0} {\hat T} {\cal L}_1^{eq}  \right]
\end{equation}
Comparing with 
\begin{equation} 
J^\mu_S= J^\mu_{can}+ J^\mu_{new}
\end{equation}
we conclude that 
\begin{equation}\label{nent}
\int d^3y \sqrt{-G}  J^0_{new} =  
\int d^3y \sqrt{-G}  \frac{{\hat T} }{T_0} {\cal L}_1^{eq}
\end{equation}
In other words the integral of $J^0_{new}$ matches with the first order 
correction to the Goldstone action. \eqref{nent} is the principal 
formal result of this subsection. It expresses a very simple relationship
between the correction to the canonical entropy current of our system and 
the first order correction to the partition function.

To what extent does \eqref{nent} determine $J^\mu_{new}$? 
The most general first order correction to the entropy current takes the 
form 
\begin{equation}\label{genent}
\begin{split}
&J^\mu_{new} = S_u u^\mu + S_\zeta \zeta^\mu + V_s^\mu\\
\end{split}
\end{equation}
where $S_u$ and $S_\zeta$ are first order scalars while $V_s^\mu$ is a  
first order vector. Notice that, to first order, 
$ X^{\mu}=S_\zeta \zeta^\mu + V_s^\mu$ is orthogonal to ${\hat u}$. It 
follows immediately from this observation that 
$$X^0 = -a_i X^i$$
Plugging this relation into \eqref{nent} we conclude that 
the contribution from $X^\mu$ to the total entropy is not Kaluza Klein 
gauge invariant
and so must vanish (see \cite{Banerjee:2012iz} for a 
discussion on related issues). 
It follows that $S_\zeta$ and $V^\mu_s$ vanish in equilibrium. 
Upto dissipative corrections, therefore, it
follows that
\begin{equation}\label{genentn}
J^\mu_{new} = S_u u^\mu                     
\end{equation}
Now comparing with \eqref{nent} it follows that 
$$\int d^3 y \sqrt{g} \left( S_u-{\cal L}^{eq}_1 \right)=0$$
so that 
\begin{equation}\label{fer}
S_u={\cal L}^{eq}_1 +{\rm total ~~ derivatives}
\end{equation}

Let us now turn to the case at hand. ${\cal L}^{eq}_1$ was listed in \eqref{ppa}. 
It is easily verified that there exist no total derivative scalars at 
one derivative order. Consequently we conclude that 
$$S_u=\frac{f_1}{\hat T}(\zeta.\partial)\hat T 
+\frac{f_2}{\hat \nu}(\zeta.\partial)\hat \nu - 
f_3 \nabla_i\left(\frac{f}{\hat T} \zeta^i\right) + {\rm dissipative}
$$
It is not difficult to verify that this expression, together with 
\eqref{genentn}, agree exactly with \eqref{ecf} in equilibrium once we 
employ the identification of parameters \eqref{idparameter}. 

In summary, the positive divergence entropy current - which we determined 
earlier in this paper - is also uniquely determined by comparison with 
the partition function for parity even superfluids at first order in the 
derivative expansion. 

\subsection{Consistency with field redefinitions}

We will now verify that the dependence of the constitutive relations 
and entropy current of the superfluid on $f_3$ is consistent with 
the transformation \eqref{shift2} of $f_3$ under the field redefinition 
\eqref{fr}. 

Recall that the stress tensor and currents of our system take the form
\begin{equation}\label{frame-1}
\begin{split}
T^{\mu\nu} =&~ (\epsilon + P) u^\mu u^\nu +  P G^{\mu\nu} +  f  \xi^\mu  \xi^\nu + \pi ^{\mu\nu}\\
J^\mu =&~  q u^\mu -  f \xi^\mu + j^\mu\\
J^\mu _s =&~  J^\mu _{can} +  J^\mu_{new} = s u^\mu - \frac{ u_\nu \pi^{\mu \nu}}{T} -  \nu  j^\mu +  J^\mu_{new}.\\
\end{split}
\end{equation}
Substituting the field redefinition \eqref{fr} into this equation 
and setting 
$$\delta \phi(x^i)= h(x^i)$$ 
(recall $h$ is a function only of space) 
we recover a new form of the stress tensor and currents  
\begin{equation}\label{frame-1}
\begin{split}
T^{\mu\nu} =&~ (\tilde\epsilon +\tilde P) u^\mu u^\nu + \tilde P G^{\mu\nu} + \tilde f \tilde \xi^\mu \tilde \xi^\nu + \tilde\pi ^{\mu\nu}\\
J^\mu =&~ \tilde q u^\mu - \tilde f \tilde \xi^\mu + \tilde j^\mu\\
J^\mu _s =&~ \tilde J^\mu _{can} + \tilde J^\mu_{new} = \tilde s u^\mu - \frac{u_\nu\tilde \pi^{\mu\nu}}{T} - \nu \tilde j^\mu + \tilde J^\mu_{new}\\
\end{split}
\end{equation}
with 
\begin{equation}\label{2minus1}
\begin{split}
\tilde \pi^{\mu\nu} &= \pi^{\mu\nu} -
\left[\frac{\partial (\epsilon +P)}{\partial\chi}u^\mu u^\nu + \frac{\partial P}{\partial\chi} G^{\mu\nu} +\frac{\partial f}{\partial\chi}\xi^\mu \xi^\nu\right] (-2 \xi.\nabla^{(4)} h)\\
&~~~~~~~~~~ ~~~~~~~~~~~~~- f(\xi^\mu G^{\nu\alpha}\nabla^{(4)}_\alpha h +\xi^\nu G^{\mu\alpha}\nabla^{(4)}_\alpha h)\\
\tilde j^\mu &= j^\mu - \left[\frac{\partial q}{\partial\chi} u^\mu- \frac{\partial f}{\partial\chi} \xi^\mu\right](-2 \xi.\nabla^{(4)} h) + fG^{\mu\alpha}\nabla^{(4)}_\alpha h\\
\tilde J^\mu_{new} &= J^\mu_{new} -
 (-2 \xi.\nabla^{(4)} h)\left(\frac{\partial s}{\partial\chi}\right)u^\mu - \frac{u_\nu (\pi^{\mu\nu} - \tilde\pi^{\mu\nu})}{T} - \nu ( j^\mu -\tilde j^\mu) \\
 &= J^\mu_{new}  +\frac{f}{T} \left(u^\mu \xi^\nu - u^\nu\xi^\mu\right)\nabla^{(4)}_\nu h =  J^\mu_{new}+\frac{Q^{\mu\nu}}{T}\nabla^{(4)}_\nu h 
\end{split}
\end{equation}
All Greek indices in \eqref{2minus1} and \eqref{frame-1} run from $1 \dots 4$ 
and are raised and lowered with the full four dimensional metric $G^{\mu\nu}$.
$\chi$ derivatives in \eqref{2minus1} are taken at fixed $T$ and 
$\nu$. In deriving last equality in  \eqref{2minus1} we have used 
the first law of thermodynamics.
$$ d\epsilon = T ds + \nu dq -\frac{f}{2}d\chi$$

We will now independently verify that our final answers for $J^\mu_{new}$ and 
the constitutive relations have this symmetry. To start with recall that, 
from \eqref{shift2} and \eqref{idparameter}, 
\begin{equation}\label{shi}
{\tilde c}_a-c_a=\frac{1}{T} \frac{\partial h}{\partial H_a}
\end{equation}
It follows immediately from \eqref{shi} that the expression for $J^\mu_{new}$  
$$J^\mu_{new}=\sum_a c_a (\partial_\nu H_a) Q^{\mu\nu}$$
(see \eqref{ecf}) transforms as predicted by the last of \eqref{2minus1}. 

We now turn to the verification that our results for transport coefficients, 
\eqref{evencond},\eqref{evencond1}, transform as predicted by \eqref{2minus1}. The 
algebra involved in a direct verification is formidable, so we will content
ourselves with an indirect check. We first recall that we have already
verified (see \eqref{logcon}) that 
\begin{equation}\label{logcons} \begin{split}
&{\mathfrak S}_a(\delta T_{\mu\nu}, \delta J_\mu)
={\mathfrak S}_a(\pi_{\mu\nu}, j_\mu)\\
&{\mathfrak V}_a(\delta T_{\mu\nu}, \delta J_\mu)
={\mathfrak V}_a(\pi_{\mu\nu}, j_\mu)\\
\end{split}
\end{equation}
in fact this equation formed the basis of one of our two methods of determining
constitutive relations. It follows that if we can show that $\delta T_{\mu\nu}$ 
and $\delta J_\mu$ obey \eqref{2minus1}, then the same will be true of 
\eqref{evencond},\eqref{evencond1}. (Recall $\delta T_{\mu\nu}$ was the first order shift 
in the stress tensor arising from first order corrections to the Goldstone 
action; $\delta J^\mu$ was similarly defined.) We will now check that 
this is indeed the case. In order to do this we first simplify the
\eqref{2minus1} specializing to the case of stationary equilibrium
\begin{equation}\label{eqshiftcurrent}
\begin{split}
( j_0 - \tilde j_0)& = -2e^{\sigma}\left[\frac{\partial}{\partial \zeta_f^2}(q + \mu f)\right](\zeta^{eq}.\partial)h\\
&= e^{\sigma} \left[\frac{\partial}{\partial \nu}\left(\frac{f}{T}\right)\right](\zeta^{eq}.\partial)h\\
( j^i - \tilde j^i) &= -f \nabla^i h- 2(\zeta^{eq}.\partial h) \frac{\partial f}{\partial\zeta_f^2}(\zeta^{eq})^i
\end{split}
\end{equation}
and 
\begin{equation}\label{eqshiftstress}
\begin{split}
(\pi_{00}- \tilde\pi_{00}) &= -2e^{2\sigma}\left[\frac{\partial (\epsilon +\mu^2 f)}{\partial \zeta_f^2}\right](\zeta^{eq}.\partial)h\\
&=2e^{2\sigma}\left[\frac{\partial}{\partial \zeta_f^2}\left(T\frac{\partial P}{\partial T} -P\right)\right](\zeta^{eq}.\partial)h\\
&=2 e^{2\sigma}\left(-\frac{ T}{2}\frac{\partial f}{\partial T} + \frac{f}{2}\right)(\zeta^{eq}.\partial)h\\
&= -T^2e^{2\sigma}\left[\frac{\partial }{\partial T}\left(\frac{f}{T}\right)\right](\zeta^{eq}.\partial)h\\\
\\
(\pi^i_0 - \tilde\pi^i_0)&= -A_0 (j^i - \tilde j^i)\\
\\
(\pi^{ij}- \tilde\pi^{ij}) &= 2(\zeta^{eq}.\partial h)\left[-\frac{f}{2} g^{ij} + \left(\frac{\partial f}{\partial \zeta_f^2}\right) (\zeta^{eq})^i (\zeta^{eq})^j\right] +f \left[(\zeta^{eq})^i\nabla^j h +(\zeta^{eq})^j\nabla^i h\right]\\
\end{split}
\end{equation}
Where each of the scalar thermodynamic functions are evaluated on the zeroth order equilibrium solution
 $$T = \hT,~~\nu = \hn,~~~(\zeta_f)_i = \zeta^{eq}_i$$
In obtaining \eqref{eqshiftcurrent} and \eqref{eqshiftstress}
$$dP =\left( \frac{\epsilon +P+\mu^2 f}{T}\right)dT + T(q + \nu f) d\nu -\frac{f}{2}d\zeta_f^2$$
In those equations all spatial indices are raised and lowered by use of 
the spatial metric $g_{ij}$ (all the free indices will run from $1$ to $3$).

We now turn to the explicit expressions for $\delta T_{\mu\nu}$ and 
$\delta J_\mu$ listed in \eqref{evencurrent}.
Substituting 
$$\tilde f_3 = f_3 +h$$
(see \eqref{shift2}) in those expressions we obtain immediate 
agreement with \eqref{eqshiftcurrent}  and 
\eqref{eqshiftstress}. This completes our verification.

\section{Constraints on parity violating constitutive relations at first order}

In this section we use the partition function 
to derive constraints on parity violating 
contributions to constitutive relations by comparison with the 
local goldstone action \eqref{poi}. As in the previous subsection, we 
find perfect agreement with the constraints obtained from the local form 
of the second law. It turns out in this case that the second law analysis 
has already been performed, in full detail, in \cite{Bhattacharya:2011tra}. We begin this 
section by reviewing the results of \cite{Bhattacharya:2011tra}, before turning to a re derivation
of those results by comparison with \eqref{poi}. 

\subsection{Review of constraints from the second law}

\subsubsection{Basis of Frame Invariants}

As we have seen above, the constitutive relations are an expansion of 
frame invariant combinations of $\pi^{\mu\nu}$ and $j^\mu$ in terms of 
independent one derivative scalars, vectors and tensors. Before even specifying
the constitutive relations, we must first specify a basis of frame invariant
expressions that we will expand in this manner. In the previous section we 
choose to work with the frame invariant scalars ${\mathfrak S}_a$ and 
frame invariant vectors ${\mathfrak V}_a$. A different choice for frame 
invariants was made in \cite{Bhattacharya:2011tra}; in order to ease comparison with the results 
of that paper, we will adapt that choice in this section. In this subsubsection
we describe the basis of frame invariants used in \cite{Bhattacharya:2011tra}.

Let 
\begin{align}\label{pios}
	\mathbf{s}_1 &= \pi^{\mu\nu} \tilde{P}_{\mu\nu} & 
	\zeta_f^2 \mathbf{s}_2 &= \zeta_f \cd \pi \cd \zeta_f \\ \notag
	\mathbf{s}_3 &= u \cd \pi \cd u & 
	\mathbf{s}_4 &= u \cd \pi \cd \zeta_f \\ \notag
	\mathbf{s}_5 &= u \cd j & 
	\mathbf{s}_6 & = \zeta_f \cd j \\ \notag
	\mathbf{s}_7 & = -\mu_{diss} && \\ \notag
	\mathbf{v}_1^{\nu} & = u_{\mu} \pi^{\mu\alpha} \tilde{P}^{\nu}_{\phantom{\nu} \alpha} & 
	\mathbf{v}_2^{\nu} & = (\zeta_f)_{\mu} \pi^{\mu\alpha} \tilde{P}^{\nu}_{\phantom{\nu} \alpha} \\ \notag
	\mathbf{v}_3^{\nu} & = \tilde{P}^{\nu}_{\phantom{\nu}\alpha}j^{\alpha} & & \\
\notag
	\mathbf{t}  
&= \tilde{P}_{\mu}^{\phantom{\mu}{\alpha}}\tilde{P}_{\nu}^{\phantom{\mu}{\beta}}\pi_{\alpha\beta} - \frac{1}{2}\tilde{P}^{\mu\nu} \tilde{P}^{\alpha\beta}\pi_{\alpha\beta}\,,&&
\end{align}
Throughout this paper $\mathbf{s}_7=-\mu_{diss}=0$ (
$\mu_{diss}$ was defined in \cite{Bhattacharya:2011eea}). However we will 
retain $s_7$ in all our formulas, in order to permit easy adaptation of 
our final results to frames in which $\mu_{diss}\neq 0$. 
\begin{equation}
	P^{\mu\nu} = G^{\mu\nu} + u^{\mu}u^{\nu} 
	\qquad
	\tilde{P}^{\mu\nu} = P^{\mu\nu} - \frac{(\zeta_f)^{\mu} (\zeta_f)^{\nu}}{(\zeta_f)^2}\,.
\end{equation}
Following \cite{Bhattacharya:2011tra} we define the row vectors 
\begin{equation} \label{sds} \begin{split}
\mathbf{s}& = \begin{pmatrix}
		\mathbf{s}_1 & \mathbf{s}_2 & \mathbf{s}_3 & 
\mathbf{s}_4 & \mathbf{s}_5 & \mathbf{s}_6 & \mathbf{s}_7 \\
              \end{pmatrix} \\
\mathbf{v}&=\begin{pmatrix} \mathbf{v}_1 & \mathbf{v}_2 & \mathbf{v}_3 
\end{pmatrix}\,.
\end{split}
\end{equation}
We also define the matrices
\begin{equation}\label{adefs}
	A^{s}=	\begin{pmatrix}
			\frac{R s}{2 q T \psi_f}\quad  & \quad\frac{B_3}{3 T}-\frac{A_3}{2 T \psi_f}\quad  &
 \quad\frac{B_2}{3 T} - \frac{A_2}{2 T \psi_f}\quad  & \quad\frac{B_1}{3 T} - \frac{A_1}{2 T \psi_f} \\
			-\frac{R s }{q \psi_f T}\quad  & 	\quad\frac{B_3}{3 T} + \frac{A_3}{T \psi_f}\quad  &
 \quad\frac{B_2}{3 T} + \frac{A_2}{T \psi_f}\quad  & \quad \frac{B_1}{3 T} + \frac{A_1}{T \psi_f}  \\
			0 & \frac{1}{T^2} & 0 & 0 \\
			-\frac{R}{T^2 \psi_f} & \frac{K_3}{T} & \frac{K_2}{T} & \frac{K_1}{T} \\
			0 & 0 & -1 & 0 \\
			-\frac{1}{T^2 \psi_f} & 0 & 0 & 0  \\
			0 & \frac{(\rho +P)K_3}{T} & \frac{(\rho +P)K_2}{T} & \frac{(\rho+P)K_1}{T} \\
		\end{pmatrix}\,,
		\quad
	A^v = \begin{pmatrix}
		-R & 0 \\
		0 & \frac{2}{T^3 \psi_f} \\
		-1 & 0
		\end{pmatrix}\,
\end{equation}
where
$$
	R=\frac{q}{\rho+P} \qquad 
	V^{\mu}=\frac{E^{\mu}}{T} - P^{\mu\nu}\partial_{\nu}\nu
$$ 
and the $A_i$'s $B_i$'s, $C_i$'s and $K_i$'s defined as follows.
\begin{equation}\label{sevdef}
\begin{split}
\nu &= \frac{\mu}{T},\quad\psi_f = \frac{\zeta_f^2}{T^2},\quad K = \frac{\nabla_\theta [ f \xi^\theta]}{\epsilon + P},\quad R= \frac{q}{\epsilon + P}\\
B_1&=-\frac{\partial}{\partial\psi_f}[\log(s)],\quad B_2=-\frac{\partial}{\partial\nu}[\log(s)],\quad B_3=-\frac{\partial}{\partial T}[\log(s)]\\
K_1&=\frac{s}{\epsilon +P}\frac{\partial}{\partial\psi_f}\left[\frac{q}{s}\right],\quad K_2=\frac{s}{\epsilon +P}\frac{\partial}{\partial\nu}\left[\frac{q}{s}\right],\quad K_3=\frac{s}{\epsilon +P}\frac{\partial}{\partial T}\left[\frac{q}{s}\right]\\
A_1&=-\frac{1}{2}- \nu\psi_f(1 - \mu R) \left[\frac{\partial}{\partial\psi_f}\left(\frac{q}{s}\right)\right] + \frac{\psi_f}{3s}\frac{\partial s}{\partial\psi_f}\\
A_2 &= - \nu\psi_f(1 - \mu R) \left[\frac{\partial}{\partial\nu}\left(\frac{q}{s}\right)\right] + \frac{\psi_f}{3s}\frac{\partial s}{\partial\nu}\\
A_3 &= -\nu\psi_f(1 - \mu R) \left[\frac{\partial}{\partial T}\left(\frac{q}{s}\right)\right] + \frac{\psi_f}{3s}\left(\frac{\partial s}{\partial T}-\frac{3 s}{T}\right)\\
V_\mu &=\frac{E_\mu}{T} -  P^\sigma_\mu \nabla_\sigma
\left[\frac{\mu}{T}\right]\\
\Omega^\mu &= \frac{1}{2}\epsilon^{\mu\nu\lambda\sigma}u_\nu \nabla_\lambda(\zeta_f)_\sigma,~~
\omega^\mu = \frac{1}{2}\epsilon^{\mu\nu\lambda\sigma}u_\nu \nabla_\lambda u_\sigma,~~
B^\mu = \frac{1}{2}\epsilon^{\mu\nu\lambda\sigma}u_\nu F_{\lambda\sigma}\,.
\end{split}
\end{equation}

In terms of \eqref{pios}-\eqref{adefs}, the frame invariant scalar, vector and 
tensor combinations of $\pi^{\mu\nu}$, $j^\mu$ and $\mu_{diss}$
are given by the row vectors
\begin{equation}\label{fic}
\mathbf{s} A^s, ~~~~\mathbf{v}_\mu A^v, ~~~~~\mathbf{t}_{\mu\nu}\,.
\end{equation}
By scalars, vectors and tensors we mean expressions which transform 
as spin 0, $\pm 1$ and $\pm 2$ representations of the $SO(2)$ symmetry that 
is left invariant by the two vectors $u^\mu$ and $\xi^\mu$ at each point in spacetime. 


\subsubsection{Constitutive Relations}

We have 4 frame invariant scalars, 2 frame invariant vectors and one 
frame invariant tensor. The most general symmetry allowed  
parity odd first derivative constitutive relations take the form
\begin{align} \label{hios}
\begin{split}
\mathbf{t}^{\mu\nu}  &= - \tilde{\eta} \tilde{\mathcal{T}}^{\mu\nu}_1 \\
\mathbf{v}^{\mu}_i A^v_{ij}  &= - \sum_{i=1}^2 \tilde{\mathcal{V}_i} \tilde{\kappa}_{ij}
- \left( \sum_{i=3}^7 \tilde{\mathcal{V}_i} \tilde{\kappa}_{ij} 
\right) \\
	\mathbf{s}_i A^s_{ij} & = -  \left( 
\sum_{i=1}^2\sum_{j=1}^4 \tilde{\mathcal{S}}_{i}\tilde{\beta}_{ij} \right)
\end{split}
\end{align}
with $\mathcal{T}$, $\tilde{\mathcal{T}}$, $\mathcal{V}$, $\tilde{\mathcal{V}}$, 
$\mathcal{S}$ and $\tilde{\mathcal{S}}$ a basis of onshell independent 
$SO(2)$ invariant tensors, vectors and scalars given in table

\begin{table}
\centering
\begin{tabular}[h]{|c|c|c|}
\hline
vector & definition & dual parity odd vector \\
 & & $\left( \epsilon^{\mu\nu\alpha\beta}u_\nu \xi_\alpha {\cal V}_\beta \right)$ \\
 & & evaluated in equilibrium \\
\hline
  ${\cal V}_1^{\mu}$ & 
  $\left( \frac{E^{\mu}}{T}-\nabla^{\mu}\left( \frac{\mu}{T}\right)\right)$ & 
  0 \\
\hline
  ${\cal V}_2^{\mu}$ & 
  $\tilde{P}^{\mu\beta}\left(\zeta_f^{\alpha}\sigma_{\alpha\beta}\right)$ & 
  0 \\
\hline
  ${\cal V}_3^{\mu}$ & 
  $\tilde{P}^{\mu\sigma} \nabla_\sigma T$ &
  $-{\hat T} V_1^i$ \\
\hline
  ${\cal V}_4^{\mu}$ &
  $\tilde{P}^{\mu\sigma} \nabla_\sigma \left(\frac{\mu}{T}\right)$ &
  $\frac{1}{T_0} V_2^i$ \\
\hline
  ${\cal V}_5^{\mu}$ &
  $\tilde{P}^{\mu\sigma} \nabla_\sigma\left(\frac{ \zeta_f^2}{T^2}\right)$ &
  $\epsilon^{ijk} V_5^i$ \\
\hline
  ${\cal V}_6^{\mu}$ &
  $\frac{\mathcal{V}_2^{c\,\mu} - \tilde{P}^{\mu\alpha}\zeta_f^{\nu}\partial_{\alpha} 
       u_{\nu}}{\zeta_f^{2}}$ & 
  $\frac{1}{2(\zeta^{eq})^2}e^{\sigma} V_4^i$ \\
\hline
  ${\cal V}_7^{\mu}$ &
  $-\frac{P^{\mu\nu} F_{\nu\alpha}\zeta_f^{\alpha}} {\zeta_f^{2}}$ &
  $-\frac{1}{(\zeta^{eq})^2} (\xi_0 V_4^i + V_3^i)$ \\
\hline
\end{tabular}
\label{feqdata}
\caption[.]{Independent fluid vector data. Here $V_m^i$ for m=1,2,3,4,5 are independent 
vectors in equilibrium defined in \eqref{oddcur}}
\end{table}
 
\begin{table}
\centering
\begin{tabular}[h]{|c|c|c|}
\hline
pseudo scalars & definition & In equilibrium \\
\hline
$\tilde{{\cal S}}_1$ & $\omega.\xi$ & -$\half e^\sigma S_2$ \\
\hline
$\tilde{{\cal S}}_2$ & $B.\xi$ & $S_1 + \xi_0 S_2$ \\
\hline
pseudo tensors & definition & In equilibrium \\
\hline
$\tilde{\mathcal{T}}_1^{\mu\nu}$ & $\str \sigma^u_{\mu\nu}$ & 0 \\
\hline
$\tilde{\mathcal{T}}_2^{\mu\nu}$ & $\str \sigma^\xi_{\mu\nu}$ & won't need  \\
\hline
\end{tabular}
\label{feqdata1}
\caption[.]{Independent fluid scalar and tensor data. Here $S_m$ for m=1,2 are independent 
vectors in equilibrium defined in \eqref{oddcur}. $\sigma^u$ and $\sigma^\xi$ are the 
shear tensors for normal and superfluid velocity respectively and 
$\str \sigma_{\mu\nu} = \epsilon^{\mu\rho\alpha\beta}u^\rho \xi^\alpha 
\sigma_{\beta}^{~\nu}+(\mu\leftrightarrow\nu)$}
\end{table}

\subsubsection{Constraints on constitutive relations from the 
local second law}

Notice that both pseudo tensors that appear in \eqref{hios} are nondissipative.
Further, the five pseudo vectors ${\cal V}_i$ $i= 3 \ldots 7$, are 
nondissipative. It may therefore come as no surprise to the reader that
\cite{Bhattacharya:2011tra} was able to use the principle of local entropy increase to determine 
${\tilde \kappa}_{im}$ ($i=3 \ldots 7$ and $m=1 \ldots 2$), together with 
${\tilde \beta}_{ij}$ ($i=1 \ldots 2$ and $j=1 \ldots 4$) in terms of 
two free functions that appeared in the parameterization of the entropy 
current. These two functions were called $\sigma_8$ and $\sigma_{10}$ in 
\cite{Bhattacharya:2011tra}. The results of \cite{Bhattacharya:2011tra} were presented in terms of $\sigma_8$ and 
$\sigma_{10}$ and four additional auxiliary fields which were determined 
in terms of $\sigma_8$ and $\sigma_{10}$ by the relations
\footnote{All terms in \eqref{arbsols} proportional to the constant ${\tilde h}$ 
were omitted in \cite{Bhattacharya:2011tra}. The reason for this is that 
\cite{Bhattacharya:2011tra} assumed that the entropy current was gauge 
invariant. As explained in \cite{Banerjee:2012iz} this does not seem 
to be physically necessary as long as the divergence of the entropy current
is gauge invariant. This allows the addition of the new term proportional to 
${\tilde h}$ in \eqref{gencuri}, which allows for a slight modification of 
the results of \cite{Bhattacharya:2011tra}, captured by the shifts 
described below. As we will see later, the requirement of CPT invariance 
forces ${\tilde h}$ to vanish.}
\begin{equation}\label{arbsols}
\begin{split}
\sigma_3&= -T \frac{\partial}{\partial T} (\sigma_{10}-\nu \sigma_8)\\
\sigma_4&= \sigma_8 + C\nu+2 {\tilde h} -\frac{\partial}{\partial \nu}  (\sigma_{10}-\nu \sigma_8)\\
\sigma_5&=-\frac{\partial}{\partial \psi_f} (\sigma_{10}-\nu \sigma_8) \\
\sigma_9&= 2\nu (\sigma_{10}-\nu \sigma_8)-\frac{2}{3}C \nu^3-2 {\tilde h}\nu^2+s_9
\end{split}
\end{equation}

In terms of all these fields, it was demonstrated in \cite{Bhattacharya:2011tra} that point wise
positivity of the the divergence of the entropy current determines
\begin{equation}\label{ll} 
{\tilde \eta}=0,~~~{\tilde \kappa}_{m2}=0
\end{equation}
and 
\begin{align}
\begin{split}
\label{solvv}
	\tilde{\kappa}_{31} & = -R T \sigma_3 - T \partial_T \sigma_8 \\
	\tilde{\kappa}_{41} & =- R T^2 \sigma_4 - T \partial_\nu \sigma_8 \\
	\tilde{\kappa}_{51} & = -R T^2 \sigma_5 - T \partial_{\psi} \sigma_8 \\
	\tilde{\kappa}_{61} & = -2 R T^3 \sigma_9 +2 T^2 \sigma_{10} \\
	\tilde{\kappa}_{71} & = -R T^2 \sigma_{10} + 2 T \sigma_8 + C T\nu+2 {\tilde h} T
\end{split}
\end{align}
{\small
\begin{equation} \label{solvs}
	-\tilde{\beta}_{ij} = \begin{pmatrix}
		\frac{2 R T \snin}{\psi_f} - \frac{2 \sten}{\psi_f} \quad&
		\quad -2\sthr - 2 T^2 K_3  \snin \quad&
		\quad -2 T \sfou -2 T^2  K_2 \snin \quad&
		\quad -2 T \sfiv - 2 K_1 T^2  \snin \\
		-\frac{C\nu+2 {\tilde h}}{T\psi_f} - \frac{2 \seig}{T \psi_f} + \frac{R \sten}{\psi_f}  \quad&
		\quad \partial_T \seig - K_3 T \sten \quad &
		\quad \partial_{\nu} \seig - K_2 T \sten \quad &
		\quad \partial_{\psi_f} \seig - K_1 T \sten
	\end{pmatrix}\,.
\end{equation}
}


\subsection{Constraints on constitutive relations from the Goldstone action}

As in the previous section, we use the Goldstone action to constrain 
transport coefficients as follows. All constraints follow from the 
analogue of \eqref{logcon} 
\begin{align} \label{hiosb}
\begin{split}
	\mathbf{t}^{\mu\nu}(\delta T_{\mu\nu}, \delta J_\mu)
  &= \mathbf{t}^{\mu\nu}(\pi_{\mu\nu}, j_\mu)\\
	\mathbf{v}^{\mu}_i A^v_{ij} (\delta T_{\mu\nu}, \delta J_\mu) &=
\mathbf{v}^{\mu}_i A^v_{ij}(\pi_{\mu\nu}, j_\mu) \\
	\mathbf{s}_i A^s_{ij}(\delta T_{\mu\nu}, \delta J_\mu) & = 
	\mathbf{s}_i A^s_{ij}(\pi_{\mu\nu}, j_\mu)
\end{split}
\end{align}
The LHS in this equation may be determined in terms of the functions 
$g_1$ and $g_2$ in the Goldstone action using \eqref{oddcur}. The RHS 
of the same equation is simplified using \eqref{hios} under the substitution
$T \rightarrow {\hat T}$, $\mu \rightarrow {\hat \mu}$, $\zeta_f \rightarrow 
\zeta^{eq}$. Under the last substitution, the parity odd first derivative 
vectors and scalars evaluate to geometric expressions. 
 Substituting these relations into the RHS of \eqref{hiosb} and equation
coefficients of independent vectors and tensors yields an expression 
for all non dissipative transport coefficients in terms of the functions 
$g_1$ and $g_2$. Using Eq.(\ref{podstress}) one obtains
\begin{eqnarray}\label{vectorod1}
 \mathbf{v}_1^{i} & =& u_{\mu} \pi^{\mu\alpha} \tilde{P}^{i}_{\phantom{\nu} \alpha} \nn\\
&=& \hT^3 (-\hn \partial_{\hT}g_1+ \partial_{\hT}g_2)V_{1}^{i}+\frac{\hT^2}{T_0}(\hn \partial_{\hn}g_1- \partial_{\hn}g_2 )V_{2}^{i}-\frac{\hT^2}{(\zeta^{eq})^2}(g_{2}-2 g_1 \hn)V_3^{i}\nn\\
&+&T_0 \hn\frac{\hT^2}{(\zeta^{eq})^2}V_{4}^i+\hT^2 (\hn \partial_{\psi_{eq}}g_1-\partial_{\psi_{eq}}g_2)V_{5}^{i}\nn\\ 
\mathbf{v}_2^{i} & =& (\zeta_f)_{\mu} \pi^{\mu\alpha} \tilde{P}^{i}_{\phantom{\nu} \alpha} =0\nn\\
\mathbf{v}_3^{i} & =& \tilde{P}^{\nu}_{\phantom{i}\alpha}j^{\alpha}\nn\\
&=&\hT (\hT \partial_{\hT}g_1~ V_1^{i}-\frac{1}{T_0}\partial_{\hn}g_1~ V_2^{i}-\frac{1}{(\zeta^{eq})^2}(2 g_1~ V_{3}^{i}+g_{2}T_0 V_{4}^{i}+(\zeta^{eq})^2 \partial_{\psi_{eq}}g_1~ V_{5}^{i}))\nn\\
\mathbf{s}_1 &=& \pi^{\mu\nu} \tilde{P}_{\mu\nu} =0\nn\\ 
\mathbf{s}_2 &=&\frac{1}{(\zeta_f)^2 } \zeta_f \cd \pi \cd \zeta_f=-\frac{2}{\hT}(\zeta^{eq})^2 \big(\partial_{\psi_{eq}}g_1~S_1+T_0 \partial_{\psi_{eq}}g_2 ~ S_2 \big)\nn\\
\mathbf{s}_3 &=& u \cd \pi \cd u=\hT(\hT\partial_{\hT}g_1-2\psi_{eq} \partial_{\psi_{eq}}g_1)S_1+\hT~T_0(\hT\partial_{\hT}g_2-2\psi_{eq} \partial_{\psi_{eq}}g_2)S_2 \nn\\ 
\mathbf{s}_4 &=& u \cd \pi \cd \zeta_f =\big(\hT^2 (g_2-2 g_1 \nn)-2(\zeta^{eq})^2 \hn \partial_{\psi_{eq}}g_1\big)S_1-S_2 T_0 \hn (g_2 \hT^2 +2 (\zeta^{eq})^2 \partial_{\psi_{eq}}g_2)\nn\\
&+&2 C_1 e^{-\sigma } S_2 T_0^3-C\frac{1}{6} A_0^2  e^{-\sigma } \left(A_0 S_2+3 S_1\right)\nn\\
\mathbf{s}_5 &=& u \cd j=-(\partial_{\hn} g_1 S_1+T_0 \partial_{\hn} g_2 S_2) \nn\\
\mathbf{s}_6 & = &\zeta_f \cd j =2\hT\big(g_1+\frac{(\zeta^{eq})^2}{\hT^2}\partial_{\psi_{eq}}g_1\big)S_1+\hT T_0\big(g_2+\frac{(\zeta^{eq})^2}{\hT^2}\partial_{\psi_{eq}}g_2\big)S_2+\frac{1}{2}C A_0 e^{-\sigma } \left(A_0 S_2+2 S_1\right)\nn\\ 
\mathbf{s}_7 & =& -\mu_{diss}=0  \nn\\ 
\end{eqnarray}
Now using Eq.(\ref{hios}) one can find out the transport coefficients ${\tilde \kappa}_{ij}$ in terms of partition function coefficients $g_1,~g_2$
as follows
\begin{eqnarray}\label{vectorod2}
{ \tilde \eta}&=&0,~~{\tilde \kappa}_{i2}=0~~for~~i\in (3~to~7)\nn\\
\kappa _{31}&=& -\frac{\hT \left((-\hn  q \hT +\epsilon + P )\partial_{\hT} g_1 +q \hT \partial_{\hT }g_2   \right)}{P+\epsilon }\nn\\
\kappa _{41}&=& -\frac{\hT \left((-\hn  q \hT+\epsilon+P )\partial_{\hn} g_1 +q \hT \partial_{\hn }g_2   \right)}{P+\epsilon }\nn\\
\kappa _{51}&=& -\frac{\hT \left((-\hn  q \hT+\epsilon+P )\partial_{\psi_{eq}} g_1 +q \hT \partial_{\psi_{eq}} g_2   \right)}{P+\epsilon }\nn\\
\kappa_{61}&=&\frac{2 \hT^2}{\epsilon+P}\left(-g_2 (-2\hn  q \hT+\epsilon+P)+2 g_1 \hn (-\hn  q \hT+\epsilon+P)\right)\nn\\
&+&\frac{C \hn ^2 \hT^2 (3 p-2 \hn  q \hT+3 \epsilon )}{3 (P+\epsilon )}-\frac{4 C_1 q \hT^3}{P+\epsilon }\nn\\
\kappa_{71}&=&\frac{ \hT}{\epsilon+P}\left(g_2   q \hT+2 g_1 (-\hn  q \hT+\epsilon+P)\right)+\frac{C \hn  \hT (2 p-\hn  q \hT+2 \epsilon )}{2 (P+\epsilon )}.\nn\\
\end{eqnarray}
Similarly, the transport coefficients $\beta_{ij}$ in terms of partition function coefficients $g_1,~g_2$
as follows
\begin{eqnarray}\label{vectorod3}
 -\beta_{11}&=&\frac{4 R \hT \hn (-g_2+g_1 \hn)}{\psi_{eq}} - \frac{2 (-g_2+2 g_1 \hn)}{\psi_{eq}}+C\frac{\hn ^2 \hT^2 (-3 P+2 \hn  q \hT-3 \epsilon )}{3 (\zeta^{eq}) ^2 (P+\epsilon )}+C_1\frac{4 q \hT^3}{(\zeta^{eq}) ^2 (P+\epsilon )}\nn\\
-\beta_{12}&=& - \frac{2 g_1}{\hT \psi_{eq}} + \frac{R (-g_2+2 g_1 \hn)}{\psi_{eq}}-C\frac{\hn  \hT (2 P-\hn  q \hT+2 \epsilon )}{2 (\zeta^{eq})^2 (P+\epsilon )} \nn\\
-\beta_{21}&=& -2 \hT (-\hn \partial_{\hT}g_1+\partial_{\hT}g_2) - 4\hn T^2 K_3  (-g_2+g_1 \hn)-\frac{2}{3}C K_3 \hT^2 \hn^3-4C_1 K_3 \hT^2\nn\\
-\beta_{22}&=&\partial_{\hT} g_1 - K_3 \hT (-g_2+2g_1\nu)-\frac{1}{2}C K_3 \hT \hn^2\nn\\
-\beta_{31}&=&-2 \hT (-\hn \partial_{\hn}g_1+\partial_{\hn}g_2) -4\hn \hT^2  K_2 (-g_2+g_1 \hn)-\frac{2}{3}C K_2 \hT^2 \hn^3-4C_1 K_2 \hT^2\nn\\
-\beta_{32}&=&\partial_{\hn} g_1 - K_2 \hT (-g_2+2 g_1 \hn)-\frac{1}{2}C K_2 \hT \hn^2\nn\\
-\beta_{41}&=&-2 T (-\hn \partial_{\psi_{eq}}g_1+\partial_{\psi_{eq}}g_2) - 4 K_1 \hn \hT^2  (-g_2+g_1 \hn)-\frac{2}{3}C K_1 \hT^2 \hn^3-4C_1 K_1 \hT^2,\nn\\
-\beta_{42}&=&\partial_{\psi_{eq}} g_1 - K_1 \hT (-g_2+2 g_1 \hn)-\frac{1}{2}C K_1 \hT \hn^2.\nn\\
\end{eqnarray}
In equations \eqref{vectorod1}, \eqref{vectorod2} and \eqref{vectorod3} the functions $g_1$, $g_2$ and all the other thermodynamics functions (like $\epsilon$, $P$, $q$ etc) as arbitrary functions of $\hT$, $\hn$ and $\psi_{eq}$.

 If 
we make the substitution
\begin{equation}
 g_{1}=\s_{8}+{\tilde h},~~g_{2}=-\s_{10}+2\hn  \s_8 +\frac{1}{2}C\hn^2+2 {\tilde h}\hn . 
\end{equation}
and introduce the auxiliary fields 
$\sigma_3$, $\sigma_4$, $\sigma_5$ and $\sigma_9$ 
which are written in terms of $\sigma_8$ and $\sigma_{10}$ in \eqref{arbsols}
then our results for nondissipative transport coefficients agree precisely\footnote{ We also need to make identification  $s_9=2 C_1,$ as will be clear below.} 
with \eqref{ll}, \eqref{solvv}, \eqref{solvs}.

\subsection{Entropy}

As in the parity even case, we may determine the parity odd contribution 
to the entropy current by a simple direct comparison with the the 
partition function. The relevant equation here is 
\begin{eqnarray}\label{podentropycur}
 W_1^{odd}+W_{anom}=  &=&\int d^3y \sqrt {- G} \left [\hn (\delta J^0_{consistent} -\delta J^0_{covariant})+J^{0}_{S\,\hbox{\tiny new}}\right]\nn\\
&=&\int d^3y \sqrt {- G} \left [\hn \delta J^0_{shift} + J^{0}_{S\,\hbox{\tiny new}}\right]
\end{eqnarray}
The term in \eqref{podentropycur} proportional to $\delta J^0_{shift}$ has 
its origin in the fact that \eqref{finstep} is correct when $J^0$ is 
taken to be the consistent $U(1)$ current. On the other hand the canonical 
entropy current of hydrodynamics is defined in terms of the covariant 
$U(1)$ current. As explained in \cite{Banerjee:2012iz} these two currents 
differ by the shift
\begin{equation}
 j^{\mu}_{shift}=\frac{C}{6}\epsilon^{\mu\nu\rho\sigma}{\cal A}_{\nu}{\cal F}_{\rho\sigma}.
\end{equation}
The contribution of this shift to the RHS of \eqref{podentropycur} evaluates 
to 
\begin{eqnarray}\label{shiftcur}
&& \int d^3y \sqrt {- G}   \hn \delta J^0_{shift}\nn\\ &=&\frac{C}{6}\int d^3y \sqrt {-G}e^{-\sigma}\hn \epsilon^{ijk}{\cal A}_{i}{\cal F}_{jk}\nn\\
&=&\frac{C}{3}\int d^3y \sqrt {g}\hn \epsilon^{ijk}\Big(A_i \partial_{j}A_k+ A_0 A_i \partial_{j}a_k -A_{i}a_j \partial_{k}A_0+A_0 a_i \partial_{j}A_k+A_0^2 a_i \partial_{j}a_k\Big)\nn\\
&=&\frac{C}{3}\int d^3y \sqrt {g}\hn \epsilon^{ijk}\Big(A_i \partial_{j}A_k+\frac{1}{2} A_0 A_i \partial_{j}a_k +\frac{3}{2}A_0 a_i \partial_{j}A_k+A_0^2 a_i \partial_{j}a_k\Big)\nn\\
&=&\frac{C}{3}\int d^3y \sqrt {g}\hn \epsilon^{ijk}\Bigg(A_i \partial_{j}A_k+\frac{1}{2} A_0 A_i \partial_{j}a_k +\frac{3}{2 \hT^2 \psi_{eq}}A_0 (a.(\zeta^{eq}) S_1-a.V_3)\nn\\
&&~~~~~~~~~~~~~~~~~~~~~~~~~~~~~~~~~~~~~~~~~~~~~~~~~~~~+\frac{A_0^2}{\hT^2 \psi_{eq}}  (a.(\zeta^{eq}) S_2-a.V_4)\Bigg)\nn\\
\end{eqnarray}
Taking this contribution to the LHS of eq.(\ref{podentropycur}) we find 
\begin{eqnarray}\label{compare1}
 &&\int d^3y \sqrt {- G}  J^{0}_{S\,\hbox{\tiny new}}\nn\\&=& W^{odd}_1+W_{anom}-\int d^3y \sqrt {- G}  \delta J^0_{shift}\nn\\
&=&\int d^3y \sqrt {g} \Bigg(g_1 S_1+ T_0 g_2 S_2+\frac{1}{\hT^2 \psi_{eq}}(C_1 T_0^2-\frac{C}{3}\hn A_0^2)  (a.(\zeta^{eq}) S_2-a.V_4)\nn\\
&&~~~~~~~~~~~~~~~~~~~~~~~~~~~~~~~~~~~~~~~~~~-\frac{1}{2 \hT^2 \psi_{eq}}C\hn A_0 (a.(\zeta^{eq}) S_1-a.V_3)\Bigg)\nn\\
\end{eqnarray}

In rest of the section we will  use \eqref{compare1} to constrain the new part of the entropy current. 
The most general form of the first order entropy current is given by
\begin{multline}\label{gencuri}
J^{\mu}_{S\,\hbox{\tiny new}} = \epsilon^{\mu\nu\rho\sigma} \partial_{\nu} \left( \s_1 T
u_{\rho} \zeta_{\sigma} \right)
+ \s_3 \tilde{\mathcal{V}}^{c\,\mu}_3
+ T \s_4 \tilde{\mathcal{V}}_4^{c\,\mu}
+ T \s_5 \tilde{\mathcal{V}}_5^{c\,\mu}\\
+ \frac{\s_8}{2} \epsilon^{\mu\nu\rho\sigma} \xi_{\nu} F_{\rho\sigma}
+T^2 \s_9 \omega^{\mu}
+T \s_{10} B^{\mu} \\
+\alpha_1 \tilde{\mathcal{V}}^{c\,\mu}_1 + \alpha_2 \tilde{\mathcal{V}}^{c\,\mu}_2
+\zeta^\mu_{f}\left[\alpha_3 (\omega\cd\zeta) + \alpha_4 (B\cd\zeta)\right]+ {\tilde h} \epsilon^{\mu\nu\lambda\sigma}{\cal A}_\nu\partial_{\lambda}{\cal A}_\sigma \\
\text{where ${\tilde h}$ is a constant}
\end{multline}
Since the first term proportional to $\s_1$ is a total derivative, it is not determined. The term proportional to $\alpha_1$ and $\alpha_2$ is also 
undetermined as $\tilde{\mathcal{V}}^{c\,\mu}_1$ and $\tilde{\mathcal{V}}^{c\,\mu}_2$ both are zero at equilibrium. We now evaluate \eqref{gencuri} in equilibrium. 
Using Table~ $1$ and Table~ $2$ and
\begin{eqnarray}
 &&\tilde{\mathcal{V}}^{c\,I}_{0}=0 ,~~~~~ \tilde{\mathcal{V}}^{c\,0\, ,I}=-a_{i} \tilde{\mathcal{V}}^{c\,i\, ,I}~~~where ~I\in (1 ~to~ 7)\nn\\
&&\omega^{0}=\frac{ e^{\sigma}}{2 (\zeta^{eq})^2}\big((a.(\zeta^{eq}))S_2-(a.V_4)\big),\nn\\
&&B^{0}=-\frac{ 1}{ (\zeta^{eq})^2}\big((a.(\zeta^{eq}))S_1-(a.V_3)\big)-\frac{A_0}{(\zeta^{eq})^2}\big((a.(\zeta^{eq}))S_2-(a.V_4)\big)\nn\\
&&\frac{1}{2} \epsilon^{\mu\nu\rho\sigma} \xi_{\nu} F_{\rho\sigma}=e^{\sigma}(S_1+ A_0 S_2)+\frac{ A_0 e^{\sigma}}{ (\zeta^{eq})^2}\big((a.(\zeta^{eq}))S_1-(a.V_3)\big)+\frac{ A_0^2 e^{\sigma}}{ (\zeta^{eq})^2}\big((a.(\zeta^{eq}))S_2-(a.V_4)\big)\nn\\
&&~~~~~~~~~~~~~~~~~~~+e^{\sigma}(a.V_2)\nn\\
&&\epsilon^{0\nu\lambda\sigma}{\cal A}_\nu\partial_{\lambda}{\cal A}_\sigma =
e^{-\sigma}\epsilon^{ijk}
\left[A_i\partial_j A_k + 2T_0\hn a_i \partial_j A_k + T_0^2\hn^2 a_i \partial_j a_k
+\partial_i\left(T_0\hn a_j A_k\right)\right]
\end{eqnarray}where $V_i$ and $S_i$ are listed in Eq.\ref{oddcur}. Now using the fact that $(\zeta^{eq})_{i}=A_i+\partial_{i}\phi$
\begin{eqnarray}
 &&\int \sqrt{-G} \epsilon^{0\nu\lambda\sigma}{\cal A}_\nu\partial_{\lambda}{\cal A}_\sigma\nn\\
&=& \int \epsilon^{ijk}
\left[A_i\partial_j A_k + 2T_0\hn a_i \partial_j A_k + T_0^2\hn^2 a_i \partial_j a_k\right]\nn\\
&=& \int \sqrt{g}\epsilon^{ijk}
\left[(\zeta^{eq})_i\partial_j (\zeta^{eq})_k + 2T_0\hn a_i \partial_j (\zeta^{eq})_k + T_0^2\hn^2 a_i \partial_j a_k\right]\nn\\
&=&\int \sqrt{g} \big(S_1+\frac{1}{(\zeta^{eq})^2} 2 T_0 \hn ((a.(\zeta^{eq}))S_1-a.V_{3})+\frac{1}{(\zeta^{eq})^2}  T_0^2 \hn^2 ((a.(\zeta^{eq}))S_2-a.V_{4})\big)\nn\\
\end{eqnarray}

 we obtain
\begin{eqnarray}\label{totentropyod}
 \int d^{3}x \sqrt{-G}J^{0}_{S\,\hbox{\tiny new}}&=& 
\int d^{3}x \sqrt{g}\Big( e^{\sigma} \hT \s_3 (a.V_1)+(\s_8-\s_4)(a.V_2)-\hT e^{\sigma} \s_5 (a.V_5)\nn\\
&+&\s_8 (S_1+ A_0 S_2)+{\tilde h} S_1-(a.(\zeta^{eq}))\big(-\frac{1}{2}\alpha_3 e^{\sigma} S_2 +\alpha_4 (S_1+ A_0 S_2)\big)\nn\\
&+& \frac{1}{(\zeta^{eq})^2}(-\hT e^{\sigma}\s_{10}+\s_8 A_0+2 {\tilde h}  T_0 \hn)\big((a.(\zeta^{eq}))S_1-(a.V_3)\big)\nn\\
&+&\frac{1}{(\zeta^{eq})^2}(\frac{e^{2\sigma}}{2}\hT^2 \s_9 -\hT e^{\sigma}\s_{10} A_0+ \s_{8} A_{0}^2+{\tilde h} T_0^2 \hn^2) \big((a.(\zeta^{eq}))S_2-(a.V_4)\big)\Big)\nn\\
\end{eqnarray}
It is convenient to introduce the following redefinitions
\begin{eqnarray}
 \s_3 &=&-\hT \partial_{\hT} X,~\s_8-\s_4 = \partial_{\hn} X+ Y,~\s_5 =- \partial_{\psi_{eq}} X+Z.
\end{eqnarray}Now using 
\begin{equation}
 \partial_{k}X=\partial_{\hT} X ~\partial_{k}\hT+\partial_{\hn} X~ \partial_{k}\hn+\partial_{\psi_{eq}} X~ \partial_{k}\psi_{eq},
\end{equation}
the first line of the Eq.\ref{totentropyod} can be rewritten as
 \begin{eqnarray}
&&\int d^{3}x \sqrt{g}  \Big( e^{\sigma} \hT \s_3 (a.V_1)+(\s_8-\s_4)(a.V_2)-\hT e^{\sigma} \s_5 (a.V_5)\Big)\nn\\
&=& \int d^{3}x \sqrt{g}\Big(T_0 \epsilon^{ijk}a_{i}(\zeta^{eq})_{j}\partial_{k}X+ Y (a.V_2)-\hT e^{\sigma} Z (a.V_5)\Big)\nn\\
&=&\int d^{3}x \sqrt{g}\Big(-T_0 X \epsilon^{ijk}(\zeta^{eq})_{i}\partial_{j}a_{k}+T_0 X \epsilon^{ijk}a_{i}\partial_{j}(\zeta^{eq})_{k}+ Y (a.V_2)-\hT e^{\sigma} Z (a.V_5)\Big)\nn\\
&=&\int d^{3}x \sqrt{g}\Big(-T_0 X S_2+T_0 X \frac{ 1}{ (\zeta^{eq})^2}\big((a.(\zeta^{eq}))S_1-(a.V_3)\big)+ Y (a.V_2)-\hT e^{\sigma} Z (a.V_5)\Big).\nn\\
\end{eqnarray}
So we obtain
\begin{eqnarray} \label{totentropyod1}
 \int d^{3}x \sqrt{-G}J^{0}_{S\,\hbox{\tiny new}}&=& 
\int d^{3}x \sqrt{g}\Big(-T_0 X S_2 + \s_8 (S_1+ A_0 S_2)+{\tilde h} S_1\nn\\
&+&(T_0 X -\hT e^{\sigma}\s_{10}+\s_8 A_0 + 2 {\tilde h} T_0 \nu) \frac{ 1}{ (\zeta^{eq})^2}\big((a.(\zeta^{eq}))S_1-(a.V_3)\big)\nn\\
&+&\frac{1}{(\zeta^{eq})^2}(\frac{e^{2\sigma}}{2}\hT^2 \s_9 -\hT e^{\sigma}\s_{10} A_0+ \s_{8} A_{0}^2+{\tilde h} T_0^2\hn^2) \big((a.(\zeta^{eq}))S_2-(a.V_4)\big)\nn\\
&-&(a.(\zeta^{eq}))\big(-\frac{1}{2}\alpha_3 e^{\sigma} S_2 +\alpha_4 (S_1+ A_0 S_2)\big)+ Y (a.V_2)-\hT e^{\sigma} Z (a.V_5)\Big).\nn\\
\end{eqnarray}
Now using \eqref{compare1} we obtain
\begin{eqnarray} \label{fre}
 &&Y=Z=0,~~\alpha_3=\alpha_4=0\nn\\
&&X= \s_{10}-\hn \s_8 -\frac{1}{2}C\hn^2-2 {\tilde h}\hn,\nn\\
&&\s_3 =-\hT \partial_{\hT} (\s_{10}-\hn \s_8),~\s_4=\s_8-\partial_{\hn} (\s_{10}-\hn \s_8)+C\hn+2{\tilde h},~\s_5=-\partial_{\psi_{eq}} (\s_{10}-\hn \s_8)\nn\\
&&\s_{9}=2\hn (\s_{10}-\hn \s_8)+2 (C_1 -\frac{C}{3}\hn^3)-2 {\tilde h} \hn^2\nn\\
&&g_{1}=\s_8+{\tilde h},~~g_{2}=-\s_{10}+2 \hn \s_{8} +\frac{1}{2}C\hn^2+2 {\tilde h}\hn.\nn\\
\end{eqnarray}
\footnote{The expression $2 C_1$ was referred to as $s_9$ in 
\cite{Bhattacharya:2011tra}.} 
It may be verified that \eqref{fre} is consistent with \eqref{arbsols}. In 
other words the entropy current determined by comparison with 
partition function agrees exactly with the non dissipative part of the 
entropy current determined from the requirement of positivity of 
divergence. \footnote{Note however that the entropy positivity method, 
in addition, determines 
two dissipative terms in the entropy current, and so, in that sense, 
carries more information about the entropy current.}

\section{CPT Invariance} \label{cpt}
In this section we explore the constraints imposed on the partition function \eqref{ppa} and \eqref{poi} by the
requirement of 4 dimensional CPT invariance. In Table $3$ we list the action of CPT on
various fields appearing in the partition function.
\begin{table}
\centering
\begin{tabular}[h]{|c|c|c|c|c|}
\hline
Field & C & P & T & CPT \\
\hline
$\sigma$ & $+$ & $+$ & $+$ & $+$ \\
\hline
$a_i$ & $+$ & $-$ & $-$ & $+$ \\
\hline
$g_{ij}$ & $+$ & $+$ & $+$ & $+$ \\
\hline
$A_0$ & $-$ & $+$ & $+$ & $-$ \\
\hline
$A_i$ & $-$ & $-$ & $-$ & $-$ \\
\hline
$\zeta_i$ & $-$ & $-$ & $-$ & $-$ \\
\hline
\end{tabular}
\caption{Action of CPT}
\label{cpt}
\end{table}
\begin{itemize}
 \item {\bf{Parity even case:}}    Using this table one easily see that, demanding CPT invariance of the action \eqref{ppa},
 the functions $f_1,~f_2,~f_3$ are even under CPT. Instead had we demanded only time reversal invariance, then the we would conclude that $f_1=f_2=f_3=0.$
 \item {\bf{Parity odd case:}} Now demanding CPT invariance of the action \eqref{poi}, we conclude that $g_1$ is odd function of $A_0$ and hence it can not 
contain any constant. This in particular implies ${\tilde h}=0,$ since $g_1=\s_8+ {\tilde h}$. So the gauge non invariant piece in entropy current in \eqref{gencuri} vanishes 
once we demand CPT invariance. The function $g_2$ appearing in \eqref{poi}  is even function in $A_0.$ 
It is also easy to see that the requirement of CPT invariance of the partition function forces $C_1 = 0.$ 

\end{itemize}

\section{Discussion}

In this paper we have studied the  equality type constraints between transport 
coefficients for relativistic superfluids at first order in the derivative 
expansion. Our central result is that the constraints obtained from 
a local form of the second law of thermodynamics agree exactly with those 
obtained from a study of the equilibrium partition function.

As the constraints obtained from both methods are numerous and rather involved in structure, 
the perfect agreement found in this paper strengthens the conjecture
 \cite{Banerjee:2012iz} that the constraints obtained from the partition 
function agree with those obtained from the local version of the second 
law of thermodynamics under all circumstances. It would be interesting 
to find either a proof for or a counterexample against this conjecture.

In the special case that the superfluid is nondissipative, \cite{Dubovsky:2011sj} has
presented a framework for describing superfluid dynamics from an action formalism. It would be interesting to understand the connection of  the
formalism of \cite{Dubovsky:2011sj} to that described in this paper.

As we have explained above, a central object in our analysis was a local 
Euclidean action for the superconducting Goldstone field. In the neighborhood 
of a second order phase transition familiar Landau-Ginzburg action for the 
order parameter is the natural analogue of the Goldstone boson action 
utilized in this paper. It seems likely that the methods of the current 
paper generalize to the study of hydrodynamics in the neighborhood 
of second order phase transitions (see \cite{RevModPhys.49.435} for a 
review). It would be interesting to perform this generalization.

Finally, in this paper we have discussed only the equality type constraints
on nondissipative transport coefficients that follow from the local second 
law. We have neither discussed Onsager type equality constraints on 
dissipative coefficients nor the inequalities on dissipative coefficients.
It is possible that these constraints follow the imposition of reasonable 
conditions (like stability) to time fluctuations about equilibrium. We 
leave the study of time dependence to future work.

\acknowledgments

We would like to thank N. Banerjee, J. Bhattacharya and A. Yarom for 
collaboration in the initial stages of this project, and several useful 
discussions. We would also like to acknowledge useful discussions with  J. Bhattacharya, K. Damle, R. Loganayagam and V. Tripathy. We would also like to 
thank N. Banerjee, J. Bhattacharya, K. Jensen, and M. Rangamani for 
comments on an earlier version of this manuscript.   
The work of S.M. was supported in part by a Swarnajayanti Fellowship.
The authors would also like to acknowledge our debt to the
people of India for their generous and steady support to research in the basic sciences.

\bibliographystyle{JHEP}
\bibliography{super}

\end{document}